\documentclass[journal]{IEEEtran}

\usepackage{framed,multirow}
\usepackage[numbers]{natbib}

\usepackage{amssymb}
\usepackage{latexsym}

\usepackage{url}
\usepackage{xcolor}
\usepackage{color}
\usepackage{amsmath,amssymb,amsfonts}
\usepackage{algorithmic}
\usepackage{graphicx}
\usepackage{textcomp}
\usepackage{booktabs}
\usepackage{bm}
\usepackage{amssymb}
\usepackage{url}
\usepackage{bbding}
\usepackage{array}
\usepackage[colorlinks,linkcolor=red]{hyperref}

\begin{document}
\title{Cross-organ all-in-one parallel compressed sensing magnetic resonance imaging}

\author{Baoshun Shi, Xin Meng, Shuangni Lv, Zheng Liu and Yan Yang
\thanks{This work was supported by the National Natural Science Foundation of China under Grants 62371414, by the Natural Science Foundation of Hebei Province under Grant F2025203070, and by the Hebei Key Laboratory Project under Grant 202250701010046. }
\thanks{B. S. Shi (Corresponding author), X. Meng, S. N. Lv and Z. Liu are with the School of Information Science and Engineering, Yanshan University, Qinhuang Dao, 066004, China. (E-mail: shibaoshun@ysu.edu.cn).\par Y. Yang (Corresponding author) is with the School of Mathematics and Statistics, Xi’an Jiaotong University, Shaanxi, 710049, China. (E-mail: yangyan@xjtu.edu.cn).}}

\maketitle

\begin{abstract}
	Recent advances in deep learning-based parallel compressed sensing magnetic resonance imaging (p-CSMRI) have significantly improved reconstruction quality. However, current p-CSMRI methods often require training separate deep neural network (DNN) for each organ due to anatomical variations, creating a barrier to developing generalized medical image reconstruction systems. To address this, we propose CAPNet (cross-organ all-in-one deep unfolding p-CSMRI network), a unified framework that implements a p-CSMRI iterative algorithm via three specialized modules: auxiliary variable module, prior module, and data consistency module. Recognizing that p-CSMRI systems often employ varying sampling ratios for different organs, resulting in organ-specific artifact patterns, we introduce an artifact generator, which extracts and integrates artifact features into the data consistency module to enhance the discriminative ability of the overall network. For the prior module, we design an organ structure-prompt generation submodule that leverages structural features extracted from the segment anything model (SAM) to create cross-organ prompts. These prompts are strategically incorporated into the prior module through an organ structure-aware Mamba submodule. Comprehensive evaluations on a cross-organ dataset confirm that CAPNet achieves state-of-the-art reconstruction performance across multiple anatomical structures using a single unified model. Our code will be published at \url{https://github.com/shibaoshun/CAPNet}.
\end{abstract}

\IEEEpeerreviewmaketitle
\section{Introduction}
	\label{Introduction}
	Magnetic resonance imaging (MRI), with its high resolution, non-invasiveness, and significant soft tissue contrast, has become widely used in the field of medical imaging. However, due to the physical characteristics of MR images, long data acquisition times are needed. 
	The time-consuming process can lead to discomfort to patients, causing inevitable motion and introducing motion artifacts in the images, which may negatively affect subsequent disease diagnosis. Therefore, accelerating MR acquisition is of significant importance in clinical practice.
Parallel imaging (PI) \citep{https://doi.org/10.1002/mrm.22161} and compressed sensing (CS) \citep{1614066} are two established acceleration strategies in MRI. PI accelerates acquisition by exploiting inter-coil encoding diversity, whereas CS leverages sparsity and incoherent undersampling to further reduce required measurements and improve reconstruction fidelity. When these two mechanisms are jointly employed, they give rise to parallel compressed sensing MRI (p-CSMRI) \citep{10.1007/978-3-030-59713-9_7}, in which multi-coil undersampled $k$-space data are acquired and reconstructed by exploiting both coil sensitivities and sparse priors within a unified framework.
While acceleration in acquisition is achieved through PI and CS strategies, reconstructing images from undersampled multi-coil data remains challenging. The reconstruction process constitutes an increasingly ill-conditioned inverse problem, which in many high-acceleration settings can become underdetermined due to insufficient independent measurements. Its stable solution requires additional prior constraints or regularization informed by the physics of the acquisition model, especially under high acceleration factors or in low-SNR scenarios. To stabilize the solution, traditional methods often incorporate sparse or other hand-crafted image priors through a regularization term $R(\cdot)$ within an optimization framework. However, traditional p-CSMRI algorithms rely on iterative optimization procedures that are computationally intensive and sensitive to the selection of regularization parameters. Their performance varies across different anatomical regions, imaging contrasts, and acceleration factors, often requiring organ-specific parameter tuning to achieve consistent reconstruction quality. These practical limitations hinder the scalability and robustness of such methods in cross-organ or high-acceleration clinical scenarios.
	
	Alternatively, deep learning (DL)-based CSMRI methods have attracted considerable attention due to their accuracy and speed \citep{SOUZA2020140, DU2021102098, 8962951, https://doi.org/10.1002/mrm.28827,10.1007/978-3-030-59713-9_7, doi:10.1137/22M1502094}. 
	These methods have achieved impressive performance in MR image reconstruction by learning robust feature representations.
Among them, deep unfolding networks bridge model-based and data-driven reconstruction by unrolling the iterations of an optimization algorithm into a structured network architecture. This design provides a more transparent and physics-aligned framework compared to end-to-end black-box models, with modules such as data consistency and prior updating corresponding to well-defined components derived from the MRI forward model. This inherent structure introduces inductive biases that can improve robustness and generalization across varying sampling patterns and anatomical regions \citep{JIANG2025113235,10926916,10453966,Jiang2023LatentspaceUF, ZHANG2025103331,10318101}.
	
	While deep unfolding networks have shown considerable promise for CSMRI, their development in academic research is still largely constrained by single-organ datasets. Major public benchmarks, such as the fastMRI brain and knee tracks \citep{Zbontar2018fastMRIAO} and the cardiac-focused CMRxRecon challenge \citep{Wang2023CMRxReconAO}, naturally promote organ-specific model training and evaluation. As a result, widely used reconstruction networks, such as MoDL \citep{8434321}, VS-Net \citep{10.1007/978-3-030-32251-9_78} and E2E-VarNet \citep{Sriram2020}, are typically developed and assessed within a single anatomical domain. Although effective in their intended settings, this practice limits scalability and motivates the design of a unified framework for robust cross-organ reconstruction. A straightforward alternative is to develop a unified deep neural network for cross-organ reconstructions. 
	Recently, all-in-one image restoration (IR) methods \citep{9879292, Valanarasu_2022_CVPR, 9157460} have gained prominence in natural image processing, aiming to use a single universal model to tackle multiple IR tasks. Can we directly apply all-in-one IR networks to cross-organ MR reconstructions? In practice, this is generally not feasible, as discussed below. These all-in-one IR methods are constrained to image-domain networks, neglecting measurement-domain knowledge. Furthermore, the inherent heterogeneity of subtle anatomical structures and tissue characteristics across different organs fundamentally limits the reconstruction quality achieved by these natural image-based all-in-one models, ultimately hindering clinical applications. These limitations highlight the need for a universal, physics-informed model capable of robustly handling cross-organ MRI reconstruction tasks within a single framework.

	In clinical practice, the optimal sampling rates vary substantially across different anatomical structures \citep{Delattre2019CompressedSM}. For cardiac imaging, subsampling strategies are routinely employed to achieve rapid acquisition, thereby mitigating motion artifacts caused by myocardial contractions \citep{XU2024105200}. Conversely, neuroimaging protocols typically require higher sampling rates to ensure sufficient spatial resolution for delineating intricate cerebral architectures.
	Moreover, different organs in clinical practice often exhibit distinct tissues and characteristics, which must be considered to achieve more accurate reconstruction. To develop a universal p-CSMRI method for multiple organs with various sampling ratios, we introduce a cross-organ all-in-one deep unfolding network for p-CSMRI in this paper. This network is capable of handling cross-organ reconstruction tasks within a single universal model.
	Specifically, we design a deep unfolding framework for p-CSMRI and implement the proposed iterative algorithm through three specialized modules: the auxiliary variable module, the prior module, and the data consistency module. For cross-organ reconstructions, useful organ-specific prompts are crucial for enhancing the network's discriminative ability. To this end, we propose an organ structure-prompt generation submodule (OSGM) that leverages structural features extracted from the segment anything model (SAM) to generate organ-specific prompts. These prompts are then integrated into the prior module through the organ structure-aware Mamba submodule (OSMM).
	Additionally, to help the model better distinguish different organs, we introduce an artifact generation submodule (AGM) that extracts image artifacts from the input and feeds these components into the network. The main contributions of this paper are as follows:
	\begin{itemize}
		
		\item We propose CAPNet, a cross-organ all-in-one deep unfolding network for p-CSMRI reconstruction that addresses the challenging problem of multi-anatomy reconstruction within a single model. CAPNet unfolds the p-CSMRI iterative algorithm into three specialized modules: an auxiliary variable module, a prior module, and a data consistency module. Our model demonstrates robust reconstruction performance across diverse anatomical structures, outperforming existing p-CSMRI and all-in-one methods.
		
		\item We develop an organ structure-prompt generation submodule that leverages structural features extracted from the SAM to generate discriminative cross-organ prompts. These prompts are strategically injected into the prior module via an elaborated organ structure-aware Mamba submodule, enabling the network to capture organ-specific structural information. Therein, this submodule enhances a residual state space model with a residual convolution to fuse global and local features.
		
		\item To reduce the organ-specific artifact patterns caused by varying sampling ratios, we design an artifact generation submodule that extracts image artifacts from the input instance. These features are integrated into the data consistency module, improving the model’s ability of distinguishing and adapting to different organs.
		
	\end{itemize}
	
	The remainder of this paper is organized as follows: Section \ref{Related work} presents the existing DL-based CSMRI algorithms and prompt learning-based all-in-one methods. Section \ref{Method} introduces the proposed parallel CSMRI algorithm and the deep unfolding network. Section \ref{Experiments} presents the experimental results. Section \ref{Discussion} provides further discussion, and Section \ref{Conclusion} concludes the paper.

	\section{Related work}
	\label{Related work}
	
	\subsection{Deep learning-based MRI algorithms}
	\label{DL-based CSMRI algorithms}
	Traditional model-based CSMRI methods typically utilize prior knowledge to enhance reconstruction performance \citep{7337391,LAI201693,10198706}. 
	However, the iterative solving process of traditional methods is often time-consuming, and the empirically designed priors may not adapt well to the varying organs.
	Compared to non-deep learning methods, DL-based CSMRI methods can achieve superior reconstruction quality and acceleration with their powerful feature representation abilities.
	The DL-based CSMRI methods usually utilize deep neural networks to learn the mapping relationship between fully sampled $k$-space data and undersampled $k$-space data, or the mapping relationship between the reconstructed MR images and the zero-filled MR images, significantly improving reconstruction speed and maintaining image quality \citep{7493320,https://doi.org/10.1049/2024/7006156,YAN2023107619,7950457,Deora_2020_CVPR_Workshops,9084142}.
	Despite the success of deep learning, further research has sought to enhance these models by embedding more robust inductive biases. For instance, \citet{mcmanus2024enforced} proposed enforced data consistency (EDC) methods to strengthen alignment with acquired $k$-space data, while \citet{fung2024l2o} introduced learning-to-optimize (L2O) paradigms to learn iterative solver updates directly from data. While these strategies enhance performance and data fidelity, deep unfolding networks provide a complementary benefit by incorporating a degree of algorithmic interpretability. In this work, we focus on a specific form of interpretability referred to as algorithmic interpretability, defined by the direct correspondence between each network component and a mathematically derived step in the underlying optimization algorithm based on the physical imaging model. This design provides a transparent computational structure that aligns with the physics of the inverse problem.
	
	Deep unfolding networks have demonstrated remarkable performance in various medical applications and have attracted widespread attention in part because their architectures provide a degree of algorithmic interpretability \citep{8434321}. By unfolding certain optimization algorithms into network architectures, they effectively integrate model-based and learning-based methods.
	In the field of CSMRI, deep unfolding methods have demonstrated excellent reconstruction performance while offering a structured, algorithmically interpretable architecture by unfolding various optimization algorithms, such as half-quadratic splitting (HQS) \citep{Jiang2023GAHQSMR}, alternating direction method of multipliers (ADMM) \citep{8550778}, iterative shrinkage-thresholding algorithm (ISTA) \citep{9428249, 9763318}, and combining them with different deep neural networks \citep{10.1007/978-3-030-59713-9_7, QIAO2025129771, QU2024107707}.
	\citet{doi:10.1137/22M1502094} unrolled the proposed proximal alternating linearized minimization algorithm into a deep neural network for parallel CSMRI and theoretically proved its convergence as an iterative algorithm. Several other representative unrolling architectures have also been developed. For instance, Sriram et al. \citep{Sriram2020} proposed an end-to-end variational network that unfolds variational optimization into a trainable architecture. Mardani et al. \citep{Mardani2018} introduced a neural proximal gradient descent framework, while Jun et al. \citep{Jun2021} developed a joint reconstruction network for simultaneous image and sensitivity estimation. More recently, Sun et al. \citep{Sun2023} presented a cross attention-based unrolling method for accelerated MRI. These approaches reflect a spectrum of design trade-offs in terms of data consistency enforcement, prior integration, and computational complexity. However, in current practice many DL-based CSMRI models are still trained and evaluated on single-organ datasets, largely following the organ-specific design of widely used public benchmarks, such as fastMRI brain and knee \citep{Zbontar2018fastMRIAO} and the cardiac-focused CMRxRecon challenge \citep{Wang2023CMRxReconAO}. This organ-wise training paradigm reduces scalability and further motivates the development of unified frameworks that can perform robust cross-organ reconstruction within a single model.

 	The scalability of deep unfolding networks is also influenced by the underlying architectural backbone. Most existing models employ convolutional neural networks (CNNs), which are computationally efficient and effective at capturing local features. However, their ability to model long-range anatomical dependencies, while in principle achievable through deeper stacking, can be less direct and efficient than architectures explicitly designed for global context modeling. While transformers overcome this limitation by capturing global context, their quadratic complexity in image resolution imposes high computational burdens for high-resolution MRI. The recently proposed selective state space models (SSMs), such as Mamba \citep{gu2024mamba}, offer a promising alternative, enabling efficient long-range dependency modeling with linear complexity. In our framework, we incorporate Mamba-based blocks into the encoder to better capture global organ-specific characteristics, thereby enhancing the adaptability of the prompt-guided deep unfolding scheme across anatomical regions and acquisition protocols.

	\subsection{Prompt learning-based all-in-one methods}
	\label{Prompt learning}
	Prompt learning techniques have seen rapid development in natural language processing and machine learning, and these methods are increasingly being applied to the field of IR. 
	Recent advancements have leveraged prompt learning to capture discriminative information across different IR tasks, enabling the "all-in-one" IR networks to handle multiple unknown degradations within a single and universal model.
	\citet{9879292} proposed an all-in-one IR network (AirNet), which can recover clean images from various degraded versions using a single network, without requiring prior knowledge of the corruption types or levels.
	Building on this, \citet{Valanarasu_2022_CVPR} developed an end-to-end transformer-based model with a novel encoder-decoder architecture that effectively identifies and removes various weather-related degradations. 
	\citet{10526271} introduced a data-driven approach that utilizes an encoder to capture features and incorporates prompts with degradation-specific information to guide the decoder in adaptively recovering images affected by different degradations.
	Similarly, \citet{10021824} introduced a patch-based IR method using denoising diffusion probabilistic models, which achieved state-of-the-art performance in both weather-specific and multi-weather IR tasks.
	Additionally, several prompt learning-based all-in-one IR methods \citep{10204072,10204275,10204770,10196308,WAN2024112324,10.5555/3666122.3669243} extract deep degradation features from degraded images as "prompts", helping the network differentiate between various IR tasks and further improving IR performance.
	
	These methods demonstrate the promise of all-in-one restoration within the domain of natural images. However, they address degradations that are fundamentally different from MRI reconstruction, which is governed by physics-based acquisition models, multi-coil encoding, and substantial anatomical variation across organs. As such, these natural-image all-in-one methods are not directly applicable as all-in-one MRI reconstruction frameworks and must be adapted when applied outside their original domain. Recently, medical researchers are exploring the potential applications of prompt learning techniques in the medical imaging field.
	For different medical image processing tasks, \citet{Yan_AllInOne_MICCAI2024} proposed a task-adaptive routing strategy that allows conflicting tasks to select different network paths in both the spatial and channel dimensions, thereby mitigating task interference.
	\citet{jatyani2025unifiedmodelcompressedsensing} proposed a unified neural operator-based model for image reconstruction in CSMRI, which is capable of adapting to various undersampling patterns and image resolutions.
	Inspired by the spirit of all-in-one methods, we consider a cross-organ MR reconstruction problem, a more clinical problem compared with single-organ parallel CSMRI problem, and propose a cross-organ all-in-one parallel CSMRI algorithm, which allows efficient reconstruction for different organs using a single model. 
	To achieve this goal, our framework synergistically integrates prompt learning with Mamba-based modules, thereby instilling the necessary context-aware adaptability.
	
	\section{Method}
	\label{Method}
	In this Section, we introduce the cross-organ all-in-one p-CSMRI algorithm in detail. 
	First, we describe the p-CSMRI optimization problem in Section \ref{Problem formulation}. 
	Secondly, we introduce the iterative stages of the proposed algorithm with three specialized steps, i.e., an auxiliary variable updating step, 
a filtering step, and a data consistency step in Section \ref{subsec:proposed_iterative_algorithm}. 
	We then build CAPNet by implementing the three proposed steps as corresponding network modules and integrating the prompt-based modules in Section \ref{Deep unfolding network}.
	Finally, we introduce the loss function of network training in Section \ref{loss function}.
	
	\subsection{Problem formulation}
	\label{Problem formulation}
	The p-CSMRI system employs multi-coil arrays, instead of single-volume coil, to reduce scanning times.
	Combining parallel imaging with compressed sensing, i.e., p-CSMRI, has become a fundamental acceleration technique in MRI.
	The sampling model of p-CSMRI can be described as the following mechanism, i.e., a linear forward imaging model for each coil index:
	\begin{equation}\label{1}
    \boldsymbol{y}_l=\boldsymbol{UF}\boldsymbol{S}_l \boldsymbol{x}+\boldsymbol{n}_l \quad \forall l \in\left\{1,2, \cdots, N_c\right\},
	\end{equation}
where $\boldsymbol{y}_l\in\mathbb{C}^{M}$ represents the undersampled $k$-space data of the $l$-th coil,
$\boldsymbol{S}_l\in\mathbb{C}^{N\times N}$ denotes the $l$-th coil sensitivity map, 
$\boldsymbol{x}\in\mathbb{C}^{N}$ represents the underlying MR image to be reconstructed,
$\boldsymbol{n}_l\in\mathbb{C}^{N}$ is the complex-value Gaussian noise in the $l$-th coil,
$N_c$ is the total number of coils,
$\boldsymbol{F}\in\mathbb{C}^{N\times N}$ is the Discrete Fourier transform matrix,
and $\boldsymbol{U}\in\mathbb{R}^{M\times N}$ is the $k$-space sampling matrix ($M\ll N$).

Given the sensitivity maps, i.e., $\boldsymbol{S}=\left\lbrace\boldsymbol{S}_1,\cdots, \boldsymbol{S}_{N_c}\right\rbrace$, 
a CSMRI optimization problem can be formulated as a regularized optimization problem:
\begin{equation}\label{2}
    \min _{\boldsymbol{x}} \frac{1}{2} \sum_{l=1}^{N_{\mathrm{c}}}
    \left\|\boldsymbol{U}\boldsymbol{F} \boldsymbol{S}_l \boldsymbol{x}-\boldsymbol{y}_l\right\|_2^2
    +\lambda R(\boldsymbol{x}),
\end{equation}
where the first term is the data fidelity term that ensures consistency with the acquired measurements. The regularization term \(R(\boldsymbol{x})\) incorporates prior knowledge about the image. From a Bayesian perspective, this formulation can be interpreted as a maximum a posteriori (MAP) estimation problem. Assuming the noise follows an i.i.d. complex Gaussian distribution, i.e., \( \boldsymbol{y}_l \mid \boldsymbol{x} \sim \mathcal{N}(\boldsymbol{UF}\boldsymbol{S}_l\boldsymbol{x}, \sigma^2\mathbf{I}) \), and the image prior is proportional to \(\exp[-\lambda R(\boldsymbol{x})]\), the MAP estimate leads directly to the objective function defined in Eqn. (\ref{2}). Thus, \(R(\cdot)\) is functionally induced by the statistical prior on \(\boldsymbol{x}\). The parameter \(\lambda\) balances the trade-off between data fidelity and regularization.

	\subsection{The proposed iterative algorithm}
	\label{subsec:proposed_iterative_algorithm}
	To solve the p-CSMRI optimization problem, as defined in Eqn. (\ref{2}), we introduce two auxiliary variables $\boldsymbol{z}\in\mathbb{C}^{N}$ and $\left\{\boldsymbol{m}_l \in \mathbb{C}^N\right\}_{l=1}^{N_{\mathrm{c}}}$. Eqn. (\ref{2}) can be rewritten into
	\begin{equation}\label{3}
		\begin{aligned}
			&\min _{\boldsymbol{z}, \boldsymbol{m}_l, \boldsymbol{x}} \frac{1}{2} \sum_{l=1}^{N_{\mathrm{c}}}\left\|\boldsymbol{U F m}_l-\boldsymbol{y}_l\right\|_2^2+\lambda R(\boldsymbol{z}) \\
			&\text { s.t. } \boldsymbol{m}_l=\boldsymbol{S}_l \boldsymbol{x}, \boldsymbol{z}=\boldsymbol{x}, \forall l \in\left\{1,2, \cdots, N_c\right\},
		\end{aligned}	
	\end{equation}
	
	Using the half quadratic splitting (HQS) algorithm, the optimization problem formulated in Eqn. (\ref{3}) can be reformulated as
	\begin{equation}\label{4}
		\min _{\boldsymbol{z}, \boldsymbol{m}_l, \boldsymbol{x}} \frac{1}{2} \sum_{l=1}^{N_{\mathrm{c}}}\left\|\boldsymbol{U F m}_l-\boldsymbol{y}_l\right\|_2^2+\lambda R(\boldsymbol{z})+  \frac{\alpha}{2} \sum_{l=1}^{N_{\mathrm{c}}}\left\|\boldsymbol{m}_l-\boldsymbol{S}_l \boldsymbol{x}\right\|_2^2+\frac{\beta}{2}\|\boldsymbol{z}-\boldsymbol{x}\|_2^2,
	\end{equation}
	where $\alpha$ and $\beta$ are the penalty parameters. The HQS algorithm has been widely adopted in CSMRI reconstruction due to its ability to decouple complex optimization problems into simpler, tractable subproblems \citep{Shi2023ReBMDual, Wang2020HQSNet, Xin2022LearnedHQS}. This property facilitates the integration of data-driven priors while preserving optimization-theoretic guarantees. In our work, HQS provides the mathematical foundation for unfolding the iterative reconstruction process into a deep network architecture. Different from previous unrolling designs such as E2EVarNet~\citep{Sriram2020}, proximal-based methods~\citep{Mardani2018}, or joint optimization approaches~\citep{Jun2021,Sun2023}, our HQS-based framework introduces an explicit auxiliary-variable branch that 
separately handles artifact modeling and structural prior learning. This decomposition enhances interpretability and facilitates cross-organ adaptation by preventing the entanglement of artifact patterns with anatomical structures. The problem defined in Eqn. (\ref{4}) is solved by alternately optimizing the auxiliary variables $\boldsymbol{z}$, $\boldsymbol{m}_l$, and the image $\boldsymbol{x}$. For the $t$-th iteration, we update these variables by using the following three steps, i.e., filtering step, auxiliary variable updating step and data consistency step. Concretely, the iteration is as follows
	\begin{equation}\label{5}
		\boldsymbol{z}^{(t)}=\underset{\boldsymbol{z}}{\arg \min }\left\{\frac{\beta}{2}\left\|\boldsymbol{z}-\boldsymbol{x}^{(t-1)}\right\|_2^2+\lambda R(\boldsymbol{z})\right\},
	\end{equation}
	\begin{equation}\label{6}
\begin{gathered}
    \boldsymbol{m}_l^{(t)} = \underset{\boldsymbol{m}_l}{\arg \min }
        \left\{\frac{1}{2} \left\| \boldsymbol{U F} \,\boldsymbol{m}_l - \boldsymbol{y}_l \right\|_2^2 + \frac{\alpha}{2} \left\| \boldsymbol{m}_l - \boldsymbol{S}_l \boldsymbol{x}^{(t-1)} \right\|_2^2\right\}, \\
    \forall l \in\{1,2,\cdots,N_c\}.
\end{gathered}
\end{equation}
	\begin{equation}\label{7}
		\boldsymbol{x}^{(t)}=\underset{\boldsymbol{x}}{\arg \min }\left\{\frac{\alpha}{2} \sum_{l=1}^{N_c}\left\|\boldsymbol{m}_l^{(t)}-\boldsymbol{S}_l\boldsymbol{x}\right\|_2^2+\frac{\beta}{2}\left\|\boldsymbol{z}^{(t)}-\boldsymbol{x}\right\|_2^2\right\} ,
	\end{equation}
	
	Filtering step: Formally, the iteration step defined in Eqn.~(\ref{5}) can be interpreted as a denoising or filtering step. While the classical HQS formulation provides an analytical solution to Eqn.~(\ref{5}), our deep unfolding framework replaces this with a learnable prior module. 
With deep unfolding networks, the filtering operation is performed by a prior network that incorporates an organ structure-aware Mamba submodule, which implicitly solves the optimization problem defined in Eqn.~(\ref{5}).

	Data consistency step: The objective function of Eqn.~(\ref{6}) consists of two simple quadratic terms with respect to $\boldsymbol{m}_l$. To derive the closed-form solution, we first obtain the image-domain formulation by setting the gradient to zero, which yields the normal equation:
	\begin{equation}\label{8}
		\left(\boldsymbol{F}^{\mathrm{H}} \boldsymbol{U}^{\mathrm{H}} \boldsymbol{U} \boldsymbol{F} + \alpha \boldsymbol{I}\right)\boldsymbol{m}_l
			= \boldsymbol{F}^{\mathrm{H}}\boldsymbol{U}^{\mathrm{H}} \boldsymbol{y}_l + \alpha \boldsymbol{S}_l \boldsymbol{x}^{(t-1)},
	\end{equation}
	The analytical solution in the image domain is then given by:
	\begin{equation}\label{9}
		\boldsymbol{m}_l = \left(\boldsymbol{F}^{\mathrm{H}} \boldsymbol{U}^{\mathrm{H}} \boldsymbol{U} \boldsymbol{F} + \alpha \boldsymbol{I}\right)^{-1}
			\left(\boldsymbol{F}^{\mathrm{H}}\boldsymbol{U}^{\mathrm{H}} \boldsymbol{y}_l + \alpha \boldsymbol{S}_l \boldsymbol{x}^{(t-1)}\right),
	\end{equation}
	For computational efficiency, we derive an equivalent $k$-space formulation. Left-multiplying both sides of Eqn.~(\ref{8}) by $\boldsymbol{F}$ and utilizing the unitary property $\boldsymbol{F}\boldsymbol{F}^{\mathrm{H}} = \boldsymbol{I}$ yields:
	\begin{equation}\label{10}
		\boldsymbol{F}\boldsymbol{m}_l^{(t)} = \left(\boldsymbol{U}^{\mathrm{H}} \boldsymbol{U} + \alpha \boldsymbol{I}\right)^{-1}
		\left(\boldsymbol{U}^{\mathrm{H}} \boldsymbol{y}_l + \alpha \boldsymbol{F} \boldsymbol{S}_l \boldsymbol{x}^{(t-1)}\right),
	\end{equation}
	where $\boldsymbol{I}$ is the identity matrix of size $ N\times N$, and $\boldsymbol{U}^{\mathrm{H}} \boldsymbol{U}$ represents a $ N\times N$ diagonal matrix, whose diagonal elements are 0 and 1. The locations of the ones in this matrix indicate that the corresponding transformation coefficients in the $k$-space are sampled. In Eqn.~(\ref{10}), $\boldsymbol{U}^{\mathrm{H}} \boldsymbol{y}_l$ represents the $k$-space measurement data obtained through zero-filled reconstruction. 
	In summary, Eqns.~(\ref{8}) and (\ref{9}) present the complete image-domain solution to the subproblem defined in Eqn.~(\ref{6}). The derivation of its mathematically equivalent $k$-space form defined in Eqn.~(\ref{10}) is crucial for practical implementation. The preference for the $k$-space formulation stems from the diagonal structure of the matrix $\boldsymbol{U}^{\mathrm{H}} \boldsymbol{U}$, which allows the inversion in Eqn.~(\ref{10}) to be computed as a simple element-wise division. This formulation provides significant computational advantages: while a general matrix inversion would require $\mathcal{O}(N^3)$ operations, the diagonal structure of $\boldsymbol{U}^{\mathrm{H}} \boldsymbol{U}$ reduces the inversion to an $\mathcal{O}(N)$ element-wise division. The overall data consistency step, dominated by the Fourier transforms, maintains $\mathcal{O}(N \log N)$ complexity, which is crucial for practical MRI reconstruction with high-dimensional data.
	
	Auxiliary variable updating step: Eqn. (\ref{7}) also contains two quadratic terms with respect to $\boldsymbol{x}$. Similarly, we take the derivative of the cost function with respect to $\boldsymbol{x}$ and set the derivative equal to zero. The closed-form solution is given by:
	\begin{equation}\label{11}
		\boldsymbol{x}^{(t)}=\left(\beta \boldsymbol{I}+\sum_{l=1}^{N_c} \alpha \boldsymbol{S}_l^{\mathrm{H}} \boldsymbol{S}_l\right)^{-1}\left(\beta \boldsymbol{z}^{(t)}+\alpha \sum_{l=1}^{N_c} \boldsymbol{S}_l^{\mathrm{H}} \boldsymbol{m}_l^{(t)}\right),
	\end{equation}
	where $\boldsymbol{S}_l^{\mathrm{H}}$ is the conjugate transpose of $\boldsymbol{S}_l$.

	So far, we have introduced all steps to iteratively solve the problem defined in Eqn.~(\ref{2}). Within the HQS framework, the filtering step serves as a regularization-driven denoising operation that incorporates prior knowledge. The data consistency step promotes consistency between the reconstruction and the measured data. It is important to note that this formulation encourages but does not guarantee exact agreement, particularly when regularization parameters are large. To address this, we introduce a learnable soft consistency mechanism where a weighting map $v \in [0,1]$, predicted from the error map, blends the current reconstruction with the measured data, allowing adaptive balance between fidelity and robustness. Finally, the auxiliary variable updating step enforces the constraints on the auxiliary variables. While this HQS-based formulation establishes a theoretical foundation, its effectiveness in cross-organ reconstruction is inherently limited by its generic design. Directly applying this iterative algorithm to cross-organ reconstruction may lead to suboptimal reconstruction quality due to the anatomical variations across different organs. To address this issue, we design an organ structure-prompt generation submodule to generate structural information, which is then incorporated into the prior step through an organ structure-aware Mamba submodule. Additionally, we design an artifact generation submodule to introduce organ-specific artifact information, caused by varying sampling rates, into the data consistency step, thereby enhancing the overall discriminative capability of the deep unfolding network.

	\begin{figure*}[!t]
		\centering
		\includegraphics[scale=0.40]{2-1.png}
		\caption{The overall structure of the cross-organ all-in-one p-CSMRI deep unfolding network (CAPNet), along with the OSGM (organ structure-prompt generation submodule) and the OSMM (organ structure-aware Mamba submodule). The network consists of $T$ stages, where each stage contains the PM (prior module), the AVM (auxiliary variable module) and the DCM (data consistency module). The initial estimated images and the structure images estimated by SAM are fed into the OSGM to generate prompts. The prompts are then input into the OSMM to guide the network training. Additionally, the initial estimated images are also fed into the AGM (artifact generation submodule) to generate artifact information, which is fed into the DCM to enrich the feature information.}
		\label{Fig. 2}
	\end{figure*}

	\subsection{The proposed deep unfolding network}
	\label{Deep unfolding network}

	The aforementioned iterative algorithm requires dozens of iterations to converge, significantly increasing computational costs and reconstruction time. To alleviate this issue, we leverage the deep unfolding technique, which provides both computational efficiency and a transparent algorithmic structure. In this framework, each network component directly corresponds to a mathematically defined operation in the underlying optimization algorithm: the Prior Module (PM) implements the proximal operation associated with the regularization subproblem, the Data Consistency Module (DCM) carries out the measurement-consistency update derived from the forward model, and the Auxiliary Variable Module (AVM) performs the variable fusion step from our optimization formulation. This correspondence provides the structural transparency that characterizes algorithmic interpretability. We unfold this iterative algorithm into a deep neural network, and design organ-aware modules for solving the cross-organ all-in-one p-CSMRI problem.  
	The cross-organ all-in-one p-CSMRI deep unfolding network (CAPNet) is shown in Fig.~\ref{Fig. 2}. Specifically, the zero-filling image $\boldsymbol{x}^{(0)}$ is taken as the input of CAPNet, and the output denoted as $\boldsymbol{x}^{(T)}$ represents the final reconstruction. Each stage of the deep unfolding network consists of three specialized modules, i.e., the PM, DCM, and AVM, corresponding to the aforementioned steps.  Building on the purely analytical HQS formulation established in Section 3.2, CAPNet unfolds the iterative process by replacing the filtering step defined in Eqn. (\ref{5}) with a learnable PM, while maintaining the analytical solutions for the data consistency and auxiliary update steps. The PM, inspired by the denoising operation in the original formulation, incorporates an organ structure-aware Mamba submodule (OSMM) to exploit structural features across spatial dimensions. These three modules are interleaved to form the network backbone.
	
	To achieve high-quality reconstructions for multiple organs, the organ structure and sampling information should be feed into the data flow of the network, improving the discriminative ability of the overall network. To this end, we introduce several organ-aware submodules: the artifact generation module (AGM) also known artifact generator captures sampling-related artifacts, the SAM module provides semantic guidance, and the organ structure-prompt generation module (OSGM) extracts structural prompts.
Since MR scanners employ different sampling ratios for various organs, the resulting artifact patterns exhibit organ-specific characteristics. These distinctive artifact patterns can be leveraged to improve both organ discrimination and reconstruction quality in cross-organ applications. The AGM processes the initial estimated image to generate artifact patterns through joint training with underlying artifact distributions. The generated artifacts are coupled to the DCM to enhance artifact reduction by influencing the auxiliary variable $\boldsymbol{m}_l$. 
The SAM foundation model produces semantic maps that serve as prompt information, guiding the network in distinguishing between different organs during reconstruction process. These semantic prompts are incorporated at each stage to enable all-in-one MR reconstruction. The OSGM further refines features by fusing information from both the estimated image and its semantic map.
	Experimental results show that the aforementioned organ-related prompts can improve the reconstruction quality. Next subsection will introduce these organ-related submodules in detail.
	
	\subsubsection{Artifact generation submodule}

\begin{figure}[!t]
		\centering
		\includegraphics[scale=0.5]{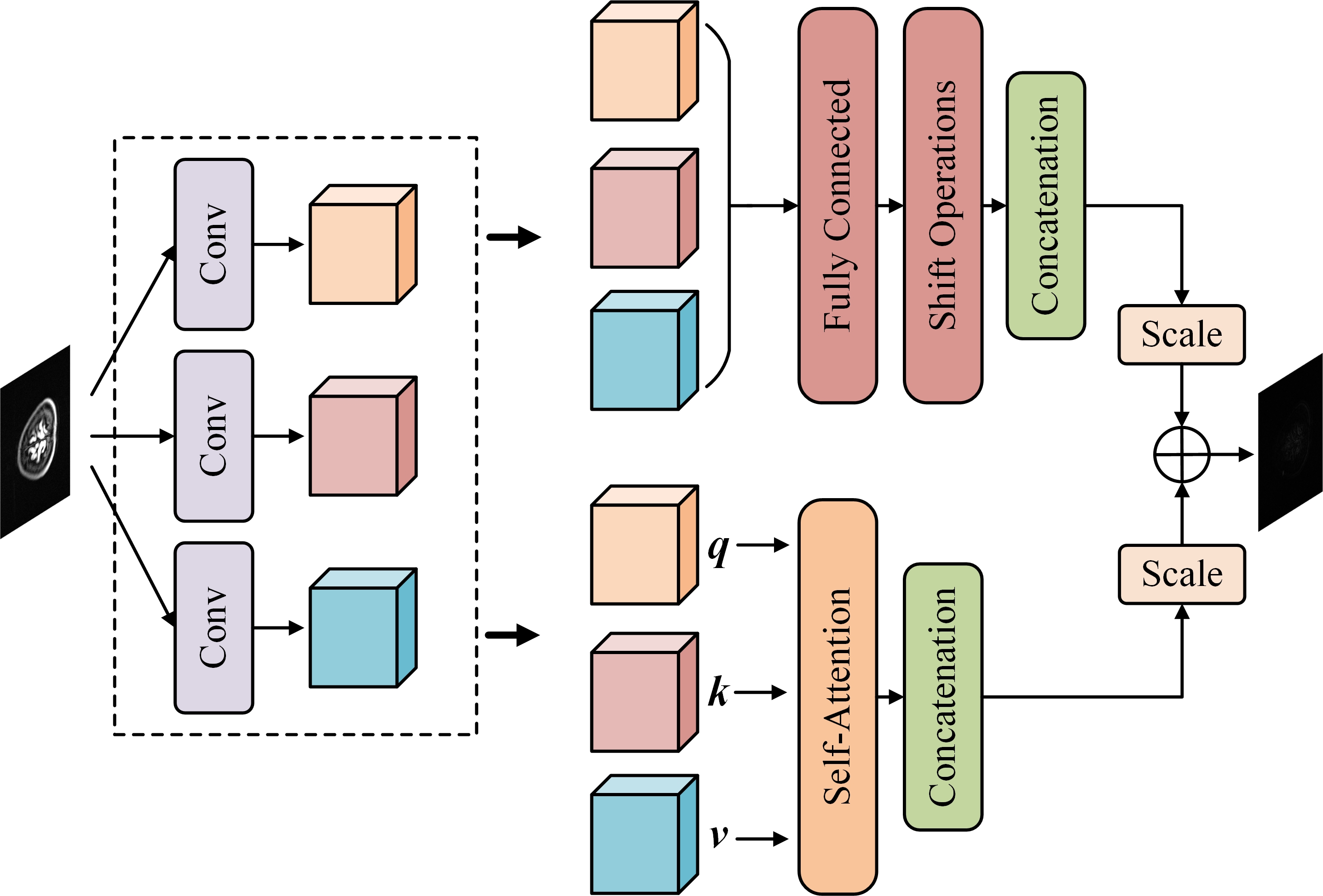}
		\caption{The structure of AGM. AGM consists of two branches: a self-attention branch and a convolutional branch.}
		\label{Fig. 3}
	\end{figure}

To achieve efficient artifact information synthesis, inspired by \citep{9878487}, we implement an artifact generation module (AGM) that synergistically integrates convolutional operations with self-attention mechanisms. As shown in Fig.~\ref{Fig. 3}, the AGM architecture employs parallel processing streams and consists of two branches: a self-attention branch and a convolutional branch.

Self-attention branch: Input features undergo dimension projection through three parallel $1\times1$ convolutional layers, generating query, key, and value matrices for multi-head attention computation.

Convolutional branch: Features are processed through channel expansion via a fully connected layer, followed by shift-based convolutional operations to extract localized spatial patterns.

The two branches exhibit complementary characteristics: the attention path excels at modeling long-range dependencies through global context aggregation, while the convolutional stream effectively preserves local structural details via shift operations. A gated fusion mechanism combines these complementary representations through element-wise summation, with their relative contributions dynamically adjusted by trainable scaling coefficients $\left( {\theta}_1, {\theta}_2 \right)\in\mathbb{R}$. This hybrid design enables simultaneous capture of both global semantic relationships and fine-grained local features, enhancing artifact generation fidelity. The above process can be formulated as
\begin{equation}\label{12}
    f_\text{artifact}(\cdot) = {\theta}_1 f_\text{att}(\cdot) + {\theta}_2 f_\text{conv}(\cdot),
\end{equation}
where $f_\text{att}(\cdot)$ denotes the self-attention branch operator, $f_\text{conv}(\cdot)$ denotes the convolutional branch operator, and $f_\text{artifact}(\cdot)$ denotes the artifact generation operator. In Eqn.~(\ref{12}), ${\theta}_1$ and ${\theta}_2$ are learnable parameters that control the strength of the two branches.

To generate diverse artifacts arising from different sampling rates and patterns, we introduce an explicit artifact supervision term that enforces consistency between the AGM output and the underlying artifact residual. In our formulation, the AGM is trained to approximate the discrepancy between an initial image estimate and the ground-truth image, so that it learns residual-based artifact representations rather than a predefined catalogue of artifact types. Under the proposed cross-organ training scheme, which covers brain, knee, and cardiac data with different acceleration factors, this residual naturally encodes aliasing from undersampling, coil-sensitivity inconsistencies, contrast variations, noise amplification, and sampling-pattern-specific distortions. As a result, the AGM learns a shared artifact manifold without requiring explicit definition of individual artifact classes. Moreover, the separation between the AGM and the organ structure-aware prior module allows artifact-related variations to be absorbed mainly in the AGM branch while preserving anatomical details in the prior branch, which improves robustness when the model encounters artifact patterns that are not explicitly represented in the training set.

	\subsubsection{Organ structure-prompt generation submodule}
	
	Since the zero-filling image contains coarse structures about different organs, we extract the semantic information of the zero-filling image by using the pre-trained SAM model, and feed both the semantic map and the zero-filling image into OSGM to generate prompt information. The detailed architecture of the OSGM is shown in Fig. \ref{Fig. 2}.
	
	The OSGM consists of two paths and is composed of different units, including one feature extraction unit and one feature fusion unit.
	First, the initial estimated image and the structure map are passed into the feature extraction unit, which consists of a $3\times 3$ convolutional layer and a rectified linear unit (ReLU) activation layer in each path to extract shallow features.  
	Then, the shallow features are input into the feature fusion unit, which comprises multiple cascaded cross-fusion residual blocks. 
	Each block consists of convolutional layers and ReLU activation layers.  
	The cross-fusion residual blocks interleave the information from two paths, aiming to fuse the rich prior information from the initial estimated image and the structure information from the structure map.  
	Finally, the cross-fused features are fed into the convolutional layers for further extraction, generating the scale denoted as $\boldsymbol{s}$ and the bias denoted as $\boldsymbol{b}$.
	The generated prompts are then input into the PM at each stage, injecting discriminative information from different organs into the data flow.  
	The procedure of OSGM can be formulated as:
	\begin{equation}\label{13}
		\left( \boldsymbol{s}, \boldsymbol{b}\right) =f_\text{OSGM}\left(\boldsymbol{x}^{(0)}, \boldsymbol{x}^{e}\right),
	\end{equation}
	where $\boldsymbol{x}^{(0)}$ is the zero-filling image, $\boldsymbol{x}^{e}$ is the semantic map generated by SAM, and $f_\text{OSGM}(\cdot)$ is the proposed organ structure-prompt generation submodule.
	
	\subsubsection{Organ structure-aware Mamba submodule}

	To fuse the prompt features extracted from the zero-filling image and its semantic map into the deep unfolding network, we build the OSMM.
The OSMM incorporates Mamba-based blocks to efficiently capture organ-specific long-range dependencies while maintaining computational efficiency. This design preserves fine anatomical structures and enforces global coherence across different organs, which is particularly benefiting for cross-organ parallel CSMRI. The module operates through three key steps:
	First, OSMM extracts global and local features from the input instances through global information extraction and local information extraction, respectively.
	Subsequently, different spatial scale features are interleaved with the prompt information from the organ structures to assist the network distinguish information from different sampling organs. 
	Finally, the features through channel shuffle operations and implements adaptive weighted feature fusion via channel attention mechanisms, thereby ensuring optimal integration and full utilization of multi-scale information.
	As the core component of the prior module, OSMM significantly improves the network's ability to distinguish between organs while enhancing overall representation capacity through its global-local feature extraction framework.
	Formally, the OSMM can be formally expressed as follows:
	\begin{equation}\label{14}
		\hat{\boldsymbol{f}}_1=\boldsymbol{s}\odot f_\text{LIE}(\text{LN}(\boldsymbol{x}^{\prime}))\oplus\boldsymbol{b},
	\end{equation}
	
	\begin{equation}\label{15}
		\hat{\boldsymbol{f}}_2=\boldsymbol{s}\odot f_\text{GIE}(\text{LN}(\boldsymbol{x}^{\prime}))\oplus\boldsymbol{b},
	\end{equation}
	
	\begin{equation}\label{16}
		\hat{\boldsymbol{f}}=f_\text{CA}\left[ f_\text{CS} \left(  \hat{\boldsymbol{f}}_1 \textcircled{c} \hat{\boldsymbol{f}}_2 \right) \right] + \boldsymbol{x}^{\prime},
	\end{equation}
	
	\begin{equation}\label{17}
		\boldsymbol{F}_\text{OSMM}=\operatorname{MLP}\left(\text{LN}\left(\hat{\boldsymbol{f}}\right)\right)+\hat{\boldsymbol{f}},
	\end{equation}
	where $\boldsymbol{x}^{\prime}$ denotes an input instance, $\textcircled{c}$ represents the concatenation of feature information across the channel dimension, $\odot$ is the element-wise product operator, $\oplus$ is the element-wise sum operator, $\operatorname{MLP}(\cdot)$ denotes multi-layer perceptron, $\operatorname{LN}(\cdot)$ refers to the layer normalization, $f_\text{LIE}(\cdot)$ indicates the local information extraction operator, $f_\text{GIE}(\cdot)$ denotes the global information extraction operator,
	$f_\text{CS}(\cdot)$ represents the channel shuffle operation, and $f_\text{CA}(\cdot)$ refers to the channel attention operator.
	
	\textbf{Global information extraction (GIE):} State space models (SSMs) have recently gained attention as a competitive paradigm to Transformers, addressing computational bottlenecks in long-sequence modeling by capturing global dependencies through linear-time operations \citep{10.1007/978-3-031-72649-1_13}. Unlike the quadratic complexity of attention mechanisms in Transformers, SSMs achieve comparable expressive power with O($n$) scaling, enabling efficient processing of extended context windows while maintaining memory efficiency. Global information reflects the overall structure, and macroscopic features of the image. SSMs are typically used to capture the global dependencies within the image to help the network understand the overall semantics of the image, thereby improving reconstruction or recognition accuracy. Building on this, we extract the global information of the input instance by using SSMs. The specific architecture of GIE is shown in Fig. \ref{Fig. 2}. We first use the LayerNorm followed by the visual state-space module (VSSM) to capture spatial long-range dependencies. Moreover, a learnable scaling factor is used to control the information from the skip connections.
	Following the VSSM module, we strategically integrate channel attention mechanisms into the global information encoder architecture to amplify cross-channel discriminative capabilities. The implementation pipeline comprises three critical phases: First, layer normalization (LayerNorm) is applied for inter-channel feature standardization, ensuring numerical stability across heterogeneous feature distributions. Subsequently, an attention-based channel gating mechanism performs adaptive feature recalibration to emphasize task-relevant channels while suppressing redundant feature components. Finally, a learnable scaling parameter modulates the residual pathway to optimize information fusion, thereby generating the refined GIE output.
	The above process can be formulated as:
	\begin{equation}\label{18}
		f_\text{VSSM}(\cdot)=\operatorname{VSSM}(\operatorname{LN}(\cdot))+c \times (\cdot),
	\end{equation}
	\begin{equation}\label{19}
		f_\text{GIE}(\cdot)=\operatorname{CA}\left(\operatorname{LN}\left(f_\text{VSSM}(\cdot)\right)\right)+c^{\prime}\times(f_\text{VSSM}(\cdot)),
	\end{equation}
	where $\operatorname{VSSM}(\cdot)$ denotes the vision state-space module, $c$ and $c^{\prime}$ are the learnable scaling factors.
	
	\textbf{Local information extraction (LIE):} Although VSSM can exploit global information, it sacrifices the inherent inductive bias of convolutions, giving rise to a relatively weak local perception ability. To compensate the VSSM, we build the local information extraction (LIE) module and integrate local information.
	While the VSSM demonstrates robust capabilities in modeling global dependencies, its architectural formulation inherently compromises the intrinsic inductive biases of convolutional operations, leading to diminished capacity for fine-grained local receptive field perception. To address this critical limitation, we design a local information extraction (LIE) as a complementary operator. The specific architecture of the LIE is shown in Fig. \ref{Fig. 2}. LIE consists of two convolutional residual blocks, where each residual block consists of two $3\times 3$ convolutional layers and a ReLU activation layer. Mathematically, the residual block is denoted as
	\begin{equation}\label{20}
		f_\text{res}(\cdot)=\operatorname{Conv}(\operatorname{ReLU}(\operatorname{Conv}(\cdot)))+(\cdot),
	\end{equation}
	where $f_\text{res}(\cdot)$ is a convolutional residual block operator, $\operatorname{ReLU}(\cdot)$ denotes the rectified linear unit activation
	layer and $\operatorname{Conv}(\cdot)$ indicates the $3\times 3$ convolutional layer.
	The process of extracting local information can be mathematically expressed as follows:
	\begin{equation}\label{21}
		f_\text{LIE}(\cdot)=f_{\text{res}}\left[f_{\text{res}}(\cdot)\right].
	\end{equation}

	\subsection{Loss function}
	\label{loss function}
	During our network training, the total loss function for network optimization consists of two parts. The primary component computes the mean square error (MSE) between the network output prediction and the ground truth image, while the secondary component is the artifact generation loss, which measures the MSE between the estimated error map and the corresponding true error map. Specifically, the total loss function is defined as
	\begin{equation}
		\mathcal{L}= \text{MSE}\left( \boldsymbol{x}_{rec}, \boldsymbol{x}_{gt}\right) + \lambda  \text{MSE}\left( \boldsymbol{x}_{este}, \boldsymbol{x}_{error}\right).
	\end{equation}
	Here, $\text{MSE}$ is the mean square error operator, $\boldsymbol{x}_{rec}$ denotes the image reconstructed by the network,
	$\boldsymbol{x}_{gt}$ signifies the ground-truth image, $\boldsymbol{x}_{este}$ represents the image artifacts generated by the AGM, and $\boldsymbol{x}_{error}$ is the underlying artifact pattern obtained by the difference between the initial estimated image and the ground-truth image.
	Heuristically, the weighting factor $\lambda$ is set as 0.1 to balance the contribution of both terms.
	Moreover, ablation experiments during model development show that MSE achieves better performance than L1 loss for artifact supervision, with average gains of about 2.1~dB in PSNR and 0.068 in SSIM. Therefore, the MSE-based loss is adopted as our final objective function.


\begin{table*}[!t]
\caption{Comparative analysis of Peak Signal-to-Noise Ratio (PSNR in dB) across different acceleration factors: Brain ($4\times$), Knee ($6\times$), and Cardiac MRI ($8\times$). Our method (CAPNet) achieves superior performance, highlighted in \textbf{bold}. $\dagger$ indicates statistically significant difference compared to CAPNet ($p < 0.001$).}
\label{tab:1}   
\centering
\setlength{\tabcolsep}{5pt}
\renewcommand{\arraystretch}{1.15}
\small
\begin{tabular}{ccccc}
\toprule 
Methods & Brain $4\times$ & Knee $6\times$ & CMR $8\times$ & Average \\
\specialrule{1pt}{0pt}{2pt} 
MoDL (\citeauthor{8434321}, \citeyear{8434321}) & $33.94 \pm 2.34^{\dagger}$ & $28.57 \pm 3.56^{\dagger}$ & $29.98 \pm 2.59^{\dagger}$ & $29.85 \pm 3.73^{\dagger}$ \\
VS-Net (\citeauthor{10.1007/978-3-030-32251-9_78}, \citeyear{10.1007/978-3-030-32251-9_78}) & $34.85 \pm 2.84^{\dagger}$ & $29.91 \pm 3.83^{\dagger}$ & $31.84 \pm 2.81^{\dagger}$ & $31.24 \pm 3.91^{\dagger}$ \\
SwinMR (\citeauthor{HUANG2022281}, \citeyear{HUANG2022281}) & $34.53 \pm 2.35^{\dagger}$ & $30.72 \pm 3.48^{\dagger}$ & $32.46 \pm 2.90^{\dagger}$ & $31.80 \pm 3.49^{\dagger}$ \\
MEDL-Net (\citeauthor{https://doi.org/10.1002/mrm.29575}, \citeyear{https://doi.org/10.1002/mrm.29575}) & $34.91 \pm 2.32^{\dagger}$ & $29.16 \pm 3.58^{\dagger}$ & $31.06 \pm 2.53^{\dagger}$ & $30.62 \pm 3.81^{\dagger}$ \\
\midrule 
PromptIR (\citeauthor{10.5555/3666122.3669243}, \citeyear{10.5555/3666122.3669243}) & $34.90 \pm 2.43^{\dagger}$ & $30.84 \pm 3.58^{\dagger}$ & $33.53 \pm 2.89^{\dagger}$ & $32.19 \pm 3.65^{\dagger}$ \\
AdaIRNet (\citeauthor{cui2024adairadaptiveallinoneimage}, \citeyear{cui2024adairadaptiveallinoneimage}) & $34.81 \pm 2.44^{\dagger}$ & $30.76 \pm 3.54^{\dagger}$ & $33.27 \pm 2.92^{\dagger}$ & $32.07 \pm 3.62^{\dagger}$ \\
CAPTNet (\citeauthor{10526271}, \citeyear{10526271}) & $34.09 \pm 2.39^{\dagger}$ & $30.60 \pm 3.52^{\dagger}$ & $32.82 \pm 2.77^{\dagger}$ & $31.74 \pm 3.48^{\dagger}$ \\
AMIR (\citeauthor{Yan_AllInOne_MICCAI2024}, \citeyear{Yan_AllInOne_MICCAI2024}) & $34.80 \pm 2.41^{\dagger}$ & $30.82 \pm 3.57^{\dagger}$ & $33.40 \pm 2.89^{\dagger}$ & $32.13 \pm 3.62^{\dagger}$ \\
\midrule 
\textbf{CAPNet} & $\bm{35.40 \pm 2.51}$ & $\bm{32.80 \pm 2.92}$ & $\bm{34.89 \pm 3.10}$ & $\bm{33.75 \pm 3.11}$ \\
\bottomrule 
\end{tabular}
\end{table*}

\begin{table*}[!t]
\caption{Comparative analysis of Structural Similarity Index (SSIM) across different acceleration factors: Brain ($4\times$), Knee ($6\times$), and Cardiac MRI ($8\times$). Our method (CAPNet) achieves superior performance, highlighted in \textbf{bold}. $\dagger$ indicates statistically significant difference compared to CAPNet ($p < 0.001$).}
\label{tab:2}   
\centering
\setlength{\tabcolsep}{5pt}
\renewcommand{\arraystretch}{1.15}
\small
\begin{tabular}{ccccc}
\toprule 
Methods & Brain $4\times$ & Knee $6\times$ & CMR $8\times$ & Average \\
\specialrule{1pt}{0pt}{2pt} 
MoDL (\citeauthor{8434321}, \citeyear{8434321}) & $0.7940 \pm 0.0735^{\dagger}$ & $0.6075 \pm 0.1738^{\dagger}$ & $0.7625 \pm 0.0889^{\dagger}$ & $0.6769 \pm 0.1660^{\dagger}$ \\
VS-Net (\citeauthor{10.1007/978-3-030-32251-9_78}, \citeyear{10.1007/978-3-030-32251-9_78}) & $0.8515 \pm 0.0564^{\dagger}$ & $0.7279 \pm 0.1573^{\dagger}$ & $0.9045 \pm 0.0292^{\dagger}$ & $0.7913 \pm 0.1461^{\dagger}$ \\
SwinMR (\citeauthor{HUANG2022281}, \citeyear{HUANG2022281}) & $0.9031 \pm 0.0351^{\dagger}$ & $0.7493 \pm 0.1503^{\dagger}$ & $0.9120 \pm 0.0274^{\dagger}$ & $0.8147 \pm 0.1407^{\dagger}$ \\
MEDL-Net (\citeauthor{https://doi.org/10.1002/mrm.29575}, \citeyear{https://doi.org/10.1002/mrm.29575}) & $0.8528 \pm 0.0542^{\dagger}$ & $0.6681 \pm 0.1556^{\dagger}$ & $0.8761 \pm 0.0383^{\dagger}$ & $0.7496 \pm 0.1570^{\dagger}$ \\
\midrule 
PromptIR (\citeauthor{10.5555/3666122.3669243}, \citeyear{10.5555/3666122.3669243}) & $0.9084 \pm 0.0364^{\dagger}$ & $0.7559 \pm 0.1530^{\dagger}$ & $0.9297 \pm 0.0240^{\dagger}$ & $0.8237 \pm 0.1441^{\dagger}$ \\
AdaIRNet (\citeauthor{cui2024adairadaptiveallinoneimage}, \citeyear{cui2024adairadaptiveallinoneimage}) & $0.9074 \pm 0.0365^{\dagger}$ & $0.7539 \pm 0.1529^{\dagger}$ & $0.9268 \pm 0.0244^{\dagger}$ & $0.8217 \pm 0.1440^{\dagger}$ \\
CAPTNet (\citeauthor{10526271}, \citeyear{10526271}) & $0.9008 \pm 0.0368^{\dagger}$ & $0.7481 \pm 0.1510^{\dagger}$ & $0.9184 \pm 0.0256^{\dagger}$ & $0.8151 \pm 0.1423^{\dagger}$ \\
AMIR (\citeauthor{Yan_AllInOne_MICCAI2024}, \citeyear{Yan_AllInOne_MICCAI2024}) & $0.9074 \pm 0.0366^{\dagger}$ & $0.7550 \pm 0.1527^{\dagger}$ & $0.9286 \pm 0.0239^{\dagger}$ & $0.8227 \pm 0.1438^{\dagger}$ \\
\midrule 
\textbf{CAPNet} & $\bm{0.8590 \pm 0.0547}$ & $\bm{0.8340 \pm 0.0913}$ & $\bm{0.9342 \pm 0.0218}$ & $\bm{0.8620 \pm 0.0851}$ \\
\bottomrule 
\end{tabular}
\end{table*}

\begin{figure}[!t]
    \centering
    \includegraphics[width=\linewidth]{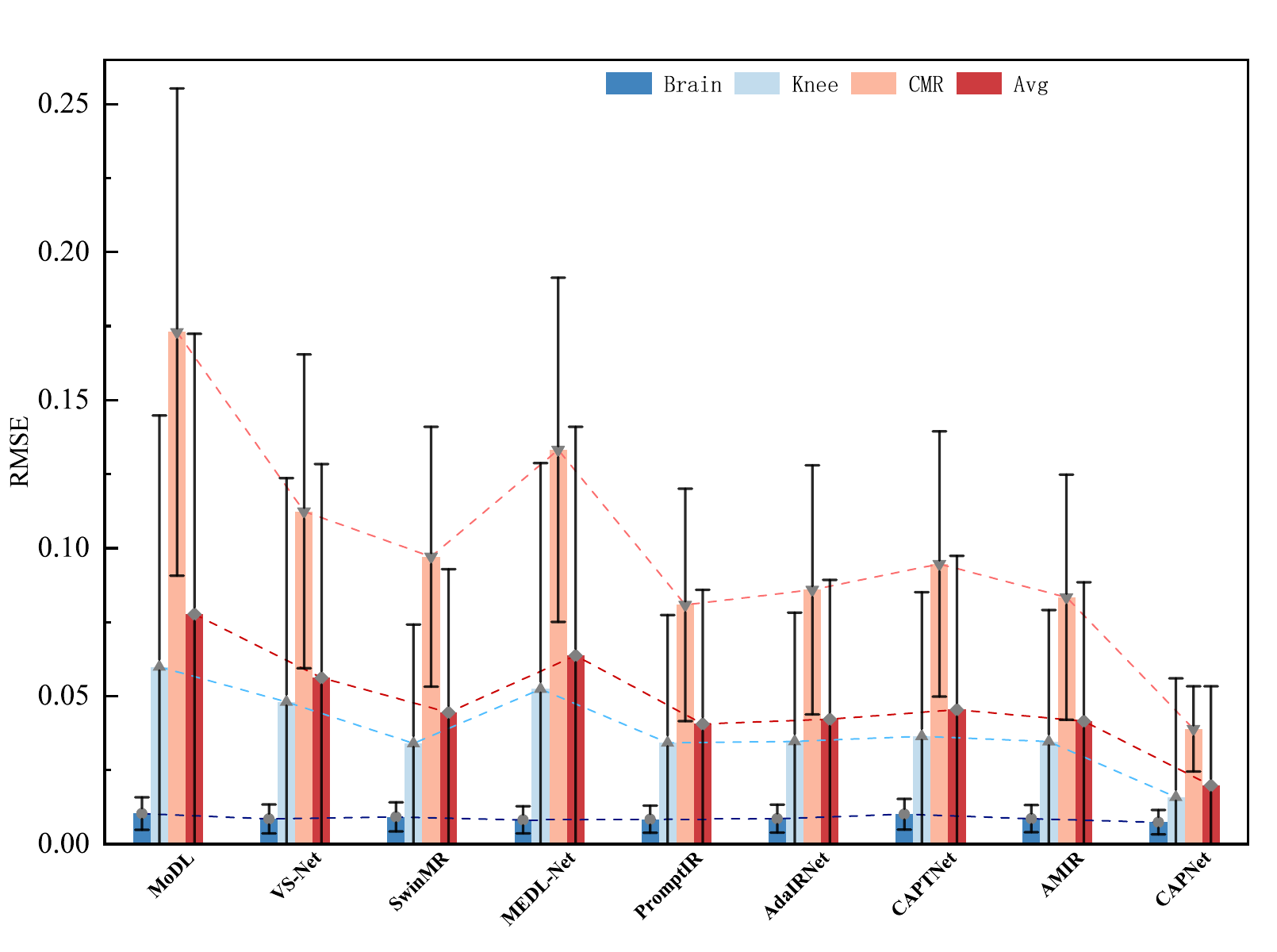}
    \caption{Root Mean Square Error (RMSE) comparison across different methods and diverse datasets. Bars represent the mean RMSE, with error bars showing the mean $\pm$ one standard deviation (truncated at zero). Colored traces connect the per-organ results for each method, facilitating cross-organ performance comparison.}
    \label{Fig. 4}
\end{figure}

\begin{figure*}[!t]
    \centering
    \includegraphics[scale=2.5]{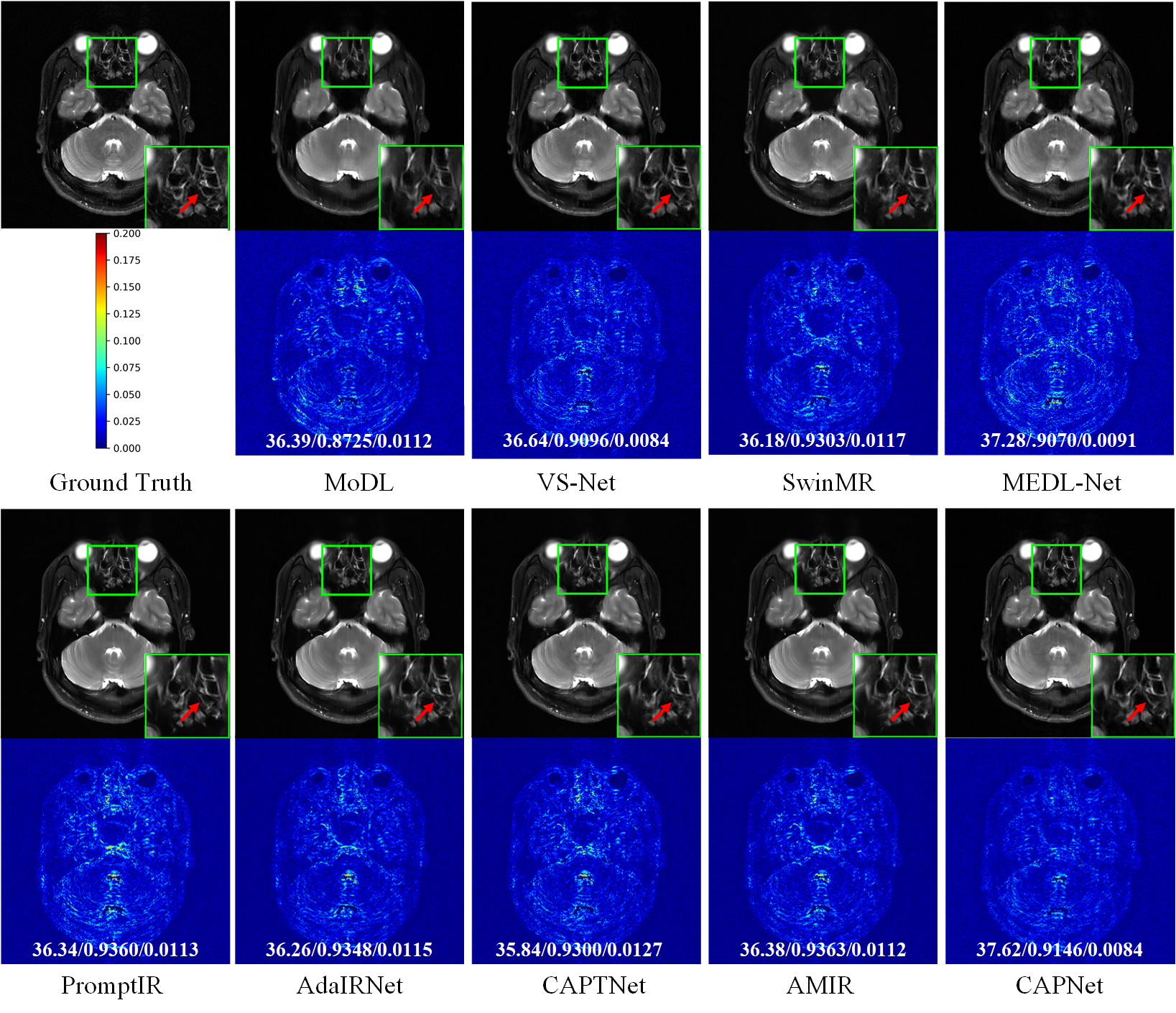}
    \caption{Visual comparison of reconstruction results and error maps for Brain MRI at $4\times$ Cartesian acceleration. Each column represents a method, with quantitative metrics (PSNR/SSIM/RMSE) annotated. The red arrow indicates regions rich in structural details.}
    \label{Fig. 5}
\end{figure*}

\begin{figure*}[!t]
    \centering
    \includegraphics[scale=2.5]{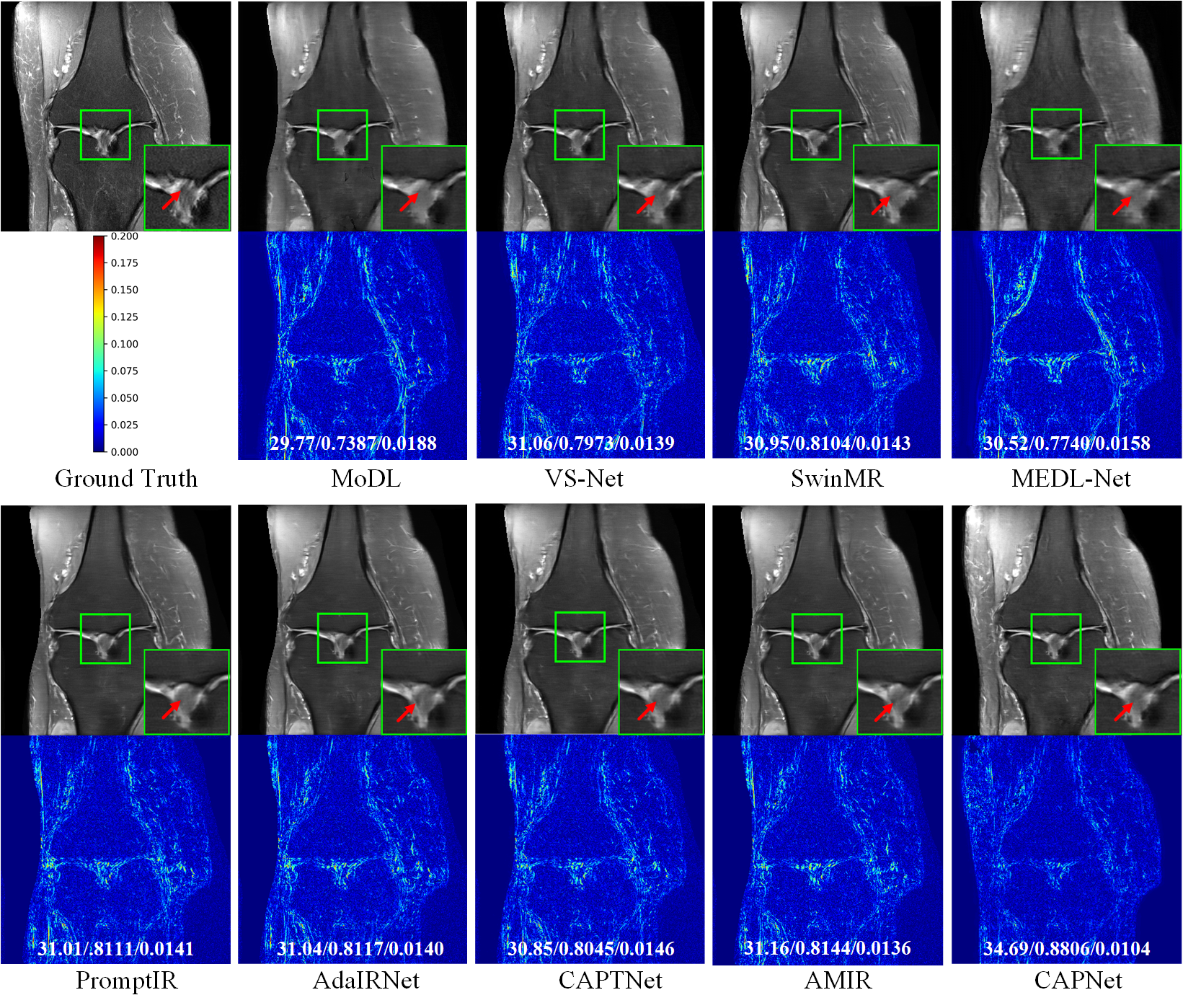}
    \caption{Visual comparison of reconstruction results and error maps for Knee MRI at $6\times$ Cartesian acceleration. Each column represents a method, with quantitative metrics (PSNR/SSIM/RMSE) annotated. The red arrow indicates regions rich in structural details.}
    \label{Fig. 6}
\end{figure*}

\begin{figure*}[!t]
    \centering
    \includegraphics[scale=2.5]{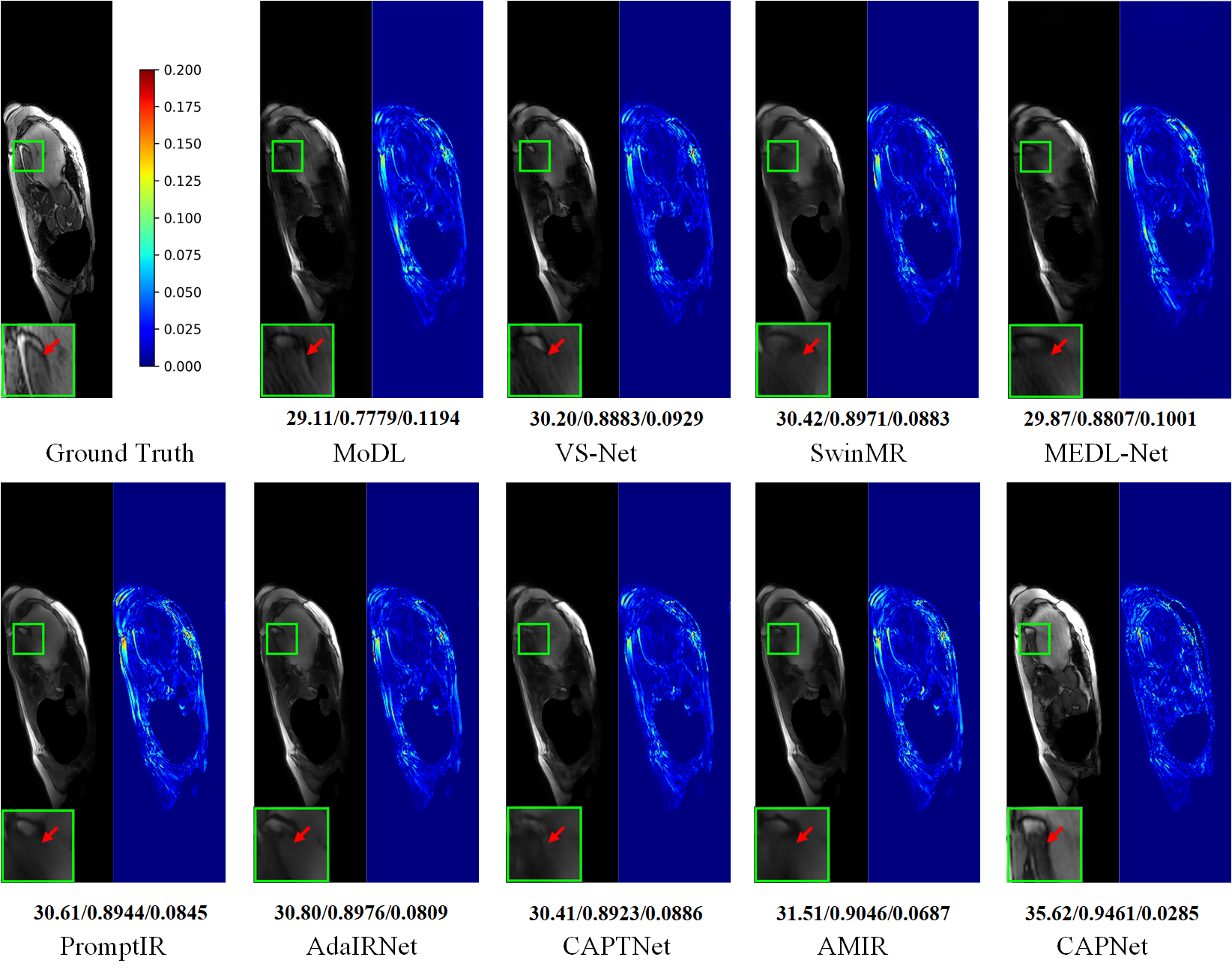}
    \caption{Visual comparison of reconstruction results and error maps for Cardiac MRI (CMR) at $8\times$ Cartesian acceleration. Each column represents a method, with quantitative metrics (PSNR/SSIM/RMSE) annotated. The red arrow indicates regions rich in structural details.}
    \label{Fig. 7}
\end{figure*}

	\section{Experiments}
	\label{Experiments}
	
	\subsection{Experiments settings}
	
	\subsubsection{Datasets}
	We use three datasets to produce a cross-organ dataset including the fastMRI dataset \citep{Zbontar2018fastMRIAO}, the CMR$\times$Recon2024 dataset \citep{Wang2023CMRxReconAO} and the in-vivo brain datasets \citep{doi:10.1137/22M1502094} to evaluate our method and the benchmark algorithms.
	The knee dataset is composed of proton density (PD) and proton density with fat saturation (PDFS) scans collected from the fastMRI database which are acquired by three clinical 3T systems (Siemens Magnetom Skyra, Prisma and Biograph mMR) or one clinical 1.5T system (Siemens Magnetom Aera), with all scans focused on the knee organ. From which, 1000/200/1000 slices (with a slice size of $15\times320\times320$) for training, validation, and testing, respectively.
	The cardiac magnetic resonance (CMR) dataset is sourced from the CMRxRecon2024 challenge and comprises T1 mapping and T2 mapping scans acquired from a 3T scanner (Siemens MAGNETOM Vida), with all imaging focused on the heart organ.
	We utilize 1000/200/400 slices for training, validation, and testing, respectively.
	The slice size uses the original size of the data.
	The brain dataset consists of T1-weighted, T2-weighted, and T1-flair raw data collected using a 3T scanner (Siemens Spectra) with a 17-channel head coil. The dataset is divided into training, validation, and testing subsets, containing 900, 150, and 300 brain slices (each of size $17\times320\times320$), respectively.
	
	\subsubsection{The benchmark methods and performance evaluation metric}
	To evaluate the performance of our CAPNet, we compare it with the state-of-the-art methods, including all-in-one methods: PromptIR \citep{10.5555/3666122.3669243}, AdaIRNet \citep{cui2024adairadaptiveallinoneimage}, CAPTNet \citep{10526271}, AMIR \citep{Yan_AllInOne_MICCAI2024}; and parallel CSMRI reconstruction methods: MoDL \citep{8434321}, VS-Net \citep{10.1007/978-3-030-32251-9_78}, SwinMR \citep{HUANG2022281}, MEDL-Net \citep{https://doi.org/10.1002/mrm.29575}.
	For the all-in-one methods,
	PromptIR uses prompts to encode degraded-specific information and then uses it to dynamically guide the restoration network.
	AdaIRNet is an adaptive all-in-one image restoration network based on frequency mining and modulation, which processes each restoration task differently by considering the impact of different degradation types on the image content in various frequency subbands.
	CAPTNet utilizes an encoder to capture features and introduces prompts with degradation-specific information to guide the decoder in adaptively restoring images affected by various degradations, enabling a single model to effectively handle multiple image degradation tasks.
	AMIR employs a task-adaptive routing strategy to address the multiple distinct medical IR tasks with a single universal model.
	For the parallel MRI reconstruction methods,
	VS-Net, MoDL and MEDL-Net are classic deep neural network architectures trained in an end-to-end manner.
	SwinMR is a Swin transformer based method for fast MRI reconstruction.
	The codes for all comparison methods were either downloaded from the authors’ repositories or implemented strictly according to the original papers. All models were trained and tested on the same datasets used in this study.
	
	To accurately and comprehensively evaluate the performance of image reconstruction, we select three metrics: Peak Signal-to-Noise Ratio (PSNR), Structural Similarity Index Measure (SSIM), and Root Mean Square Error (RMSE). These metrics are employed to quantify the similarity between the reconstructed images and their corresponding ground truth. It is noteworthy that higher PSNR and SSIM values, coupled with lower RMSE values, indicate superior reconstruction performance of the model.

\begin{table}[!t]
\caption{Average results across Brain $4\times$, Knee $6\times$, and CMR $8\times$: 
PSNR (dB)$\uparrow$ / SSIM$\uparrow$ / NMSE$\downarrow$. 
Our method (CAPNet) is highlighted in \textbf{bold}.}
\label{tab:3}   
\centering
\setlength{\tabcolsep}{5pt}
\renewcommand{\arraystretch}{1.15}
\small
\begin{tabular}{cc}
\toprule 
\multirow{2}{*}{\centering Methods} & Average \\
\cmidrule[0.7pt](lr){2-2} 
 & PSNR / SSIM / RMSE \\
\specialrule{1pt}{0pt}{2pt} 
w/o GIE & $31.02 \pm 3.86$ / $0.7839 \pm 0.1455$ / $0.0582 \pm 0.0741$ \\
w/o LIE & $31.26 \pm 3.90$ / $0.7892 \pm 0.1446$ / $0.0556 \pm 0.0712$ \\
w/o OSGM & $31.00 \pm 3.80$ / $0.7855 \pm 0.1453$ / $0.0587 \pm 0.0739$ \\
w/o AGM & $31.42 \pm 3.98$ / $0.7912 \pm 0.1458$ / $0.0539 \pm 0.0713$ \\
\midrule 
\textbf{CAPNet} & $\bm{33.75 \pm 3.11 / 0.8620 \pm 0.0851 / 0.0198 \pm 0.0335}$ \\
\bottomrule 
\end{tabular}
\end{table}

\begin{figure}[!t]
	\centering
	\includegraphics[scale=1.2]{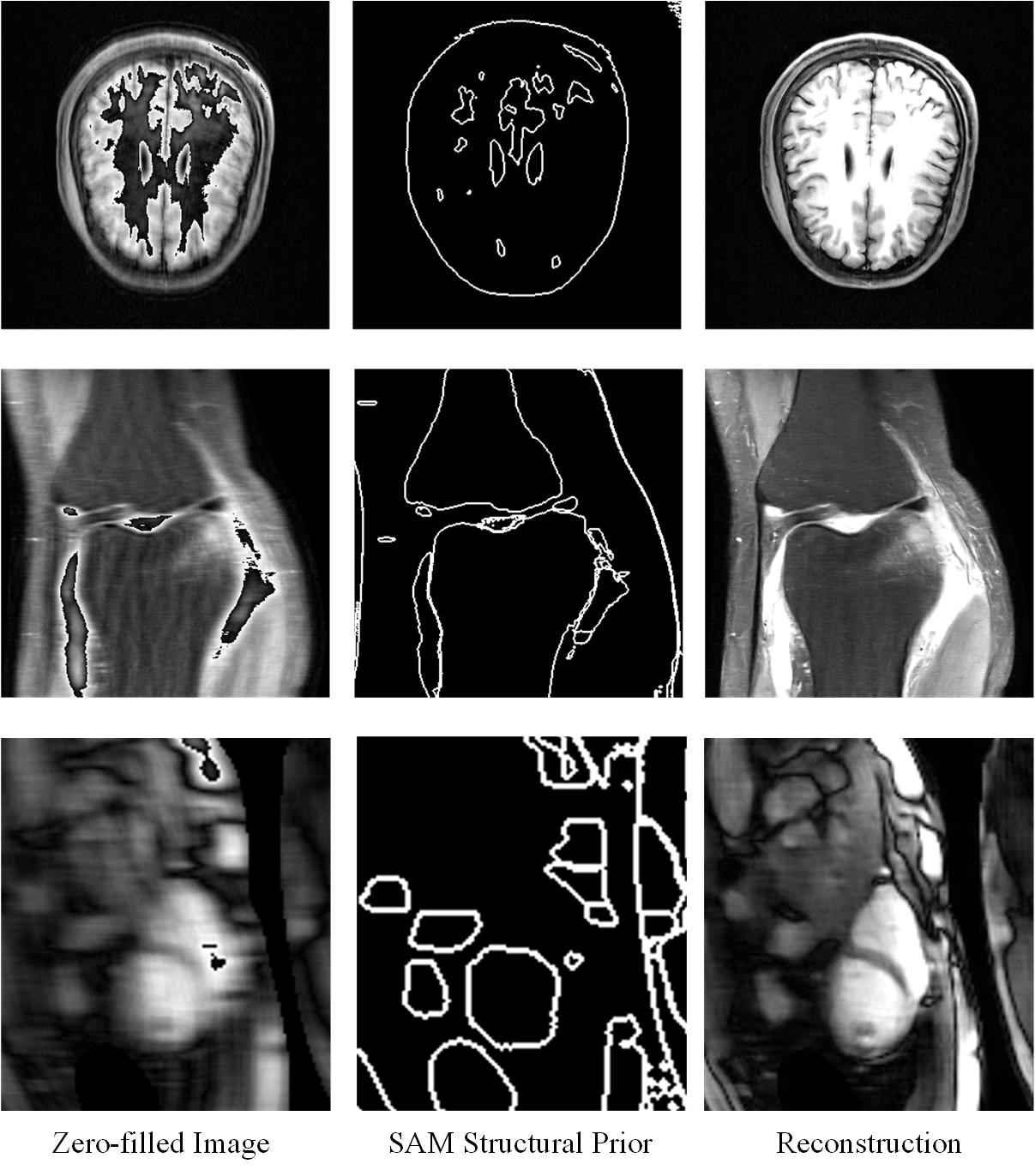}
	\caption{The CAPNet reconstruction pipeline, which leverages an auxiliary structural guidance from SAM. Illustrated from left to right: the input image also known zero-filled image, the auxiliary prompts extracted by SAM and the output of CAPNet.}	
		\label{Fig. 8}
	\end{figure}

	\subsubsection{Implementation details}
	In our experiments, we follow the paradigm set by the clinical standards \citep{Delattre2019CompressedSM}, where under-sampled measurements from three different organs are obtained by masking the corresponding fully-sampled data in $k$-space using a Cartesian sampling pattern.
	Specifically, we employ three different sampling rates to thoroughly evaluate and compare the performance of CAPNet with other methods for all-in-one reconstruction of different organs.
	For selecting the sampling rate, in actual MRI acquisition, the rate is typically adjusted based on the physician's experience and the specific clinical situation. Referring to relevant studies, we propose the following formulation:
	\begin{itemize}
		\item Since the brain area is relatively homogeneous and less affected by motion artifacts, an acceleration factor of $R{=}4$ is adopted. This setting aligns with established fastMRI benchmarks \citep{Zbontar2018fastMRIAO, Muckley2021fastMRIChallenge, Sriram2020} and represents a scenario where deep learning methods provide substantial performance improvements beyond what conventional parallel imaging methods can achieve.
    
		\item For knee imaging, which requires high spatial resolution to capture fine details, an acceleration factor of $R{=}6$ is employed. This clinically relevant factor, validated in studies such as VS-Net \citep{10.1007/978-3-030-32251-9_78}, provides a meaningful challenge that clearly distinguishes advanced reconstruction techniques from those effective only at lower accelerations.
    
		\item For cardiac imaging, which demands high speed and minimal motion artifacts, an acceleration factor of $R{=}8$ is selected. This choice is grounded in dynamic CMR studies \citep{Giese2014TR3Dflow, Neuhaus2019CS4Dflow, Peper2020PSCS4Dflow} and represents a high-acceleration scenario where the advantages of deep learning-based reconstruction algorithms are most pronounced and clinically necessary.
	\end{itemize}
	
	To generate the undersampled data at these acceleration factors, 1D Cartesian masks were applied along the phase-encoding direction using organ-specific patterns consistent with established practices: random Cartesian sampling for knee data \citep{Zbontar2018fastMRIAO}, equispaced sampling with a random offset for brain data \citep{Muckley2021fastMRIChallenge}, and clinically adopted patterns for cardiac data \citep{Delattre2019CompressedSM}.
	
	Coil sensitivity maps provided in the dataset are precomputed from a data block of size $24 \times 24$ at the center of fully sampled $k$-space using ESPIRiT. 
	Although ESPIRiT can yield multiple sets of sensitivity maps, the fastMRI preprocessing pipeline retains only the dominant eigenvector set (largest eigenvalue), providing one map per coil. Following established benchmark practices \citep{8434321, Sriram2020, doi:10.1137/22M1502094}, we utilize this single normalized set of coil sensitivity maps in our reconstruction model.  To train a unified model under the all-in-one setting, we combine all three aforementioned datasets to produce a cross-organ dataset and train a single model that is later evaluated on multiple tasks.

	All network models used in these experiments are implemented using the PyTorch framework and are run on a computer equipped with a NVIDIA GeForce GTX 3090 GPU. During the model training phase, we set the batch size to be 1, adopt an end-to-end training strategy, and choose a model size with $stage=3$ for our experiments. All parameters are optimized using the Adam optimizer with a learning rate of $2e^{-4}$. To prevent overfitting, we used a validation set to select the optimal parameters during the training process. The final experimental results are based on these optimal parameters, as selected and validated through the validation set, and applied on the test set.

	\begin{figure*}[!t]
		\centering
		\includegraphics[scale=0.5]{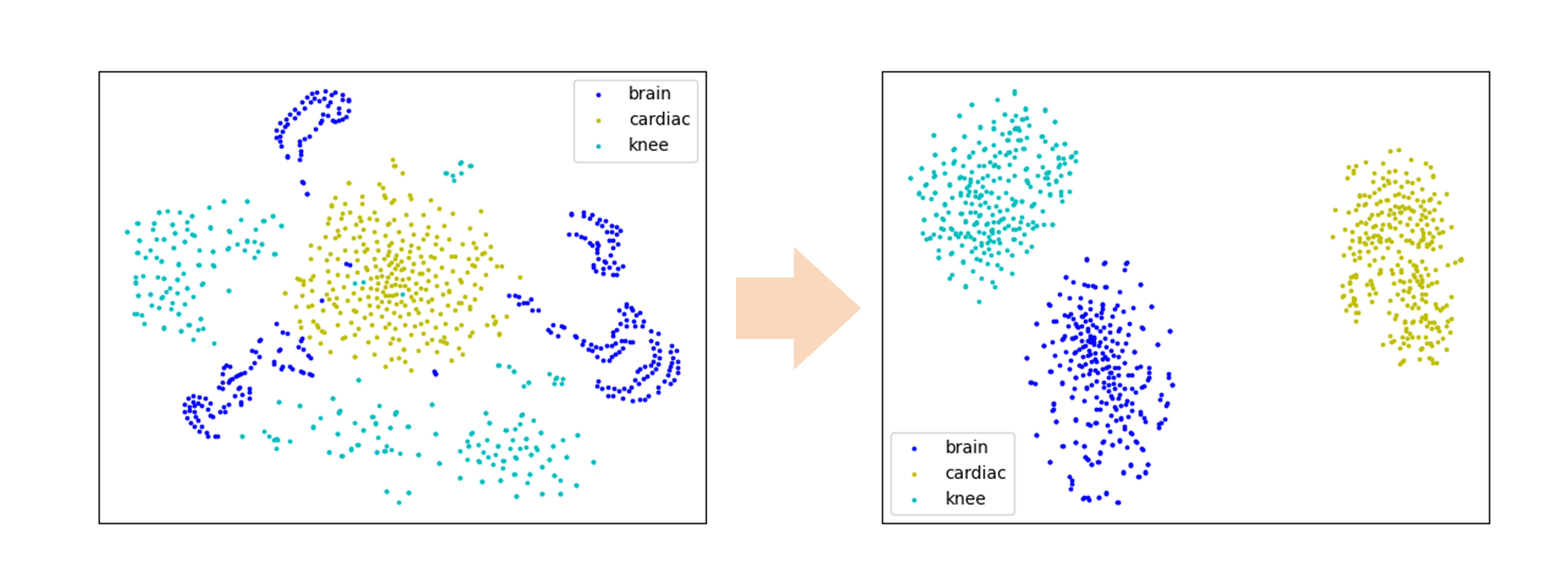}
		\caption{t-SNE visualization of feature distributions across different organs, denoted by distinct colors. (Left) Feature clustering prior to OSMM processing, demonstrating mixed organ representations. (Right) Enhanced feature separation after OSMM processing, revealing clear organ-specific clustering. These results demonstrate that the organ structure-prompt and structure-aware mechanisms effectively improve inter-organ feature discriminability in the learned representations.}
		\label{Fig. 9}
	\end{figure*}
	
\begin{figure}[!t]
		\centering
		\includegraphics[scale=0.35]{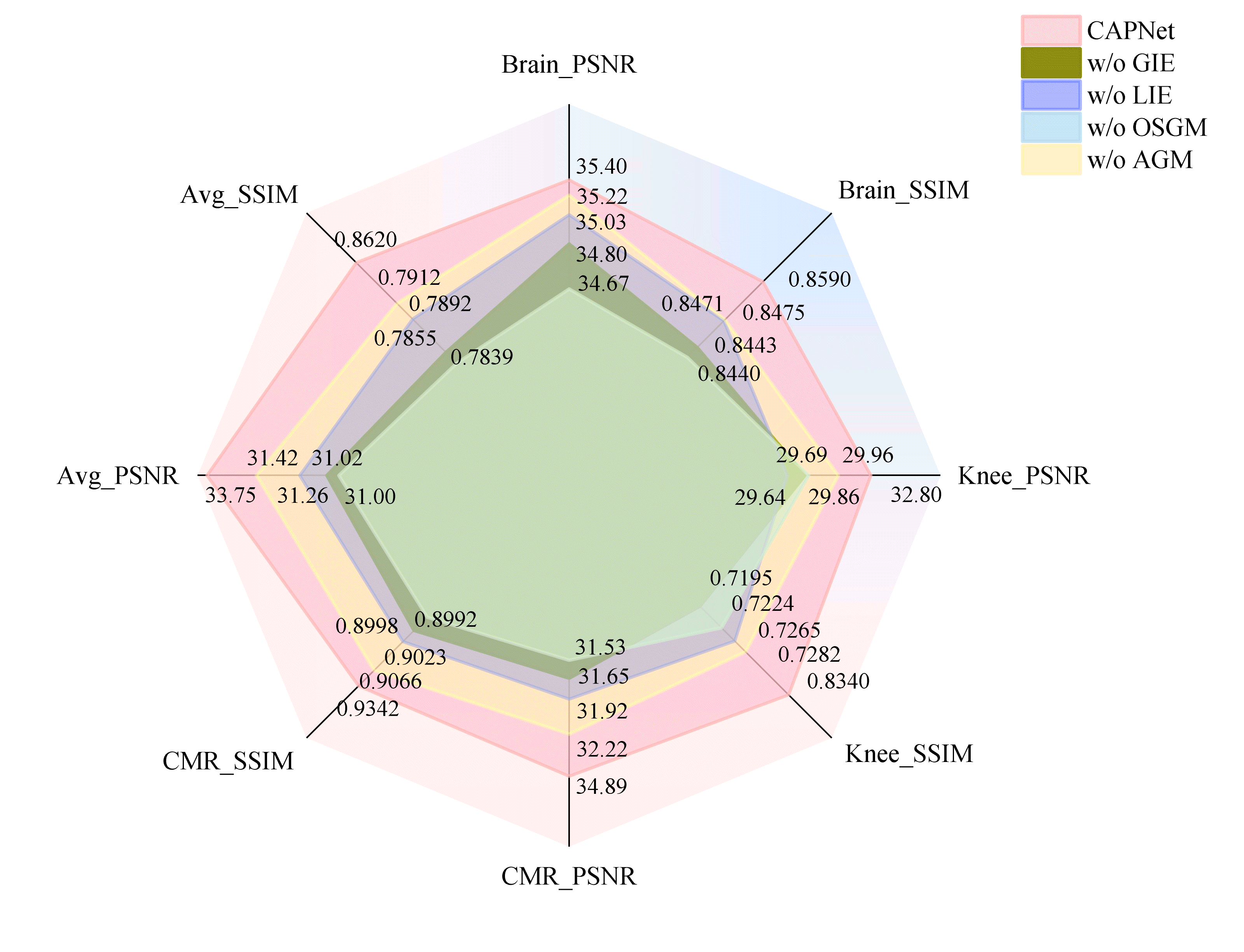}
		\caption{Ablation study results presented via radar charts, evaluating performance across key metrics for each anatomy and their averages. The chart compares the complete CAPNet model against its ablated variants (w/o GIE, w/o LIE, w/o OSGM, w/o AGM). A larger polygon area indicates better overall performance.}
		\label{Fig. 10}
	\end{figure}	

	\subsection{Experiments results}
	
	We compare CAPNet with four all-in-one methods (PromptIR, AdaIRNet, CAPTNet, AMIR) and four p-CSMRI reconstruction algorithms (MoDL, VS-Net, SwinMR, MEDL-Net) on a cross-organ dataset consisting of data from three different organs.
	Tables~\ref{tab:1}--\ref{tab:2} report the quantitative comparison results. When averaged across different reconstruction tasks, CAPNet leads all comparison methods in PSNR, SSIM and RMSE. Compared with the parallel CSMRI methods, its average PSNR value is improved by about 1.95-3.90 dB and RMSE value is reduced by 0.0247-0.0579; compared with the all-in-one method, the PSNR value is improved by 1.56-2.01 dB and the SSIM value is improved by 0.0383-0.0469. 
	
	Specifically, the proposed CAPNet significantly outperforms other methods in brain data, with a PSNR value of 35.40 dB, an improvement of 0.49 dB and 0.50 dB over the optimal benchmark p-CSMRI method MEDL-Net and the all-in-one method PromptIR, respectively. Meanwhile, its RMSE value is reduced compared to the comparison methods by 0.0008-0.0029, indicating significant advantages in detail recovery and error control.
	In the knee task, CAPNet achieves the highest values of PSNR and SSIM, which are further improved compared to SwinMR and AMIR, which verifies the robustness of CAPNet to complex structure reconstruction.
	CAPNet demonstrates excellent performance in cardiac MRI reconstruction, with PSNR and SSIM values significantly outperforming all comparison methods. Its RMSE value is 0.0419 lower than that of the next best method, indicating a unique advantage in motion artifact suppression. The results demonstrate that CAPNet is able to adaptively handle complex MRI reconstruction tasks with multiple organs by prompt learning technique, while avoiding the problem of dependence on specific-organ reconstruction problems in p-CSMRI methods.
These quantitative findings are further corroborated by the comprehensive RMSE analysis presented in Fig.~\ref{Fig. 4}, which displays the mean and standard deviation of RMSE values for brain (\emph{R}=4), knee (\emph{R}=6), and cardiac (\emph{R}=8) datasets, along with the cross-organ average. CAPNet consistently achieves the lowest RMSE values across all evaluation scenarios, with the error bars indicating its stable performance. To statistically validate the observed improvements, we performed paired t-tests and Wilcoxon signed-rank tests on a per-volume basis between CAPNet and each baseline method. All comparisons demonstrated statistically significant differences ($p < 0.001$), indicating that the performance gains achieved by CAPNet are consistent and not attributable to random variation.
	
	Fig.~\ref{Fig. 5} illustrates a qualitative comparison of the reconstructed images and their corresponding error maps using various methods under a Cartesian sampling mask for brain data from the cross-organ dataset. Compared to other methods, our approach reconstructs images with the least visible artifacts and discrepancies relative to the ground truth. Specifically, our method accurately restores intricate details in the zoomed-in regions of interest, such as subtle tissue boundaries and small anatomical features, which are either blurred or entirely missing in the results of other methods. This enhanced reconstruction quality demonstrates the effectiveness of our approach in MRI reconstructions.
	
	Similar to the brain dataset, the knee and cardiac MR datasets also demonstrate the superior performance of our method. For the knee data, as shown in Fig.~\ref{Fig. 6}, our approach effectively reduces artifacts and preserves fine structural details, such as cartilage and ligaments, which are often blurred or lost in p-CSMRI and all-in-one methods. The error maps further highlight the minimal reconstruction discrepancies achieved by our method. For the cardiac MR data, as shown in Fig.~\ref{Fig. 7}, our method successfully restores intricate vascular structures and tissue boundaries, outperforming both p-CSMRI and all-in-one methods in terms of clarity and detail preservation. These results collectively underscore the robustness and generalizability of our approach across diverse organs. 

Figs.~\ref{Fig. 5}--\ref{Fig. 7} further show that CAPNet preserves high-frequency anatomical structures more reliably than the comparison methods. The zoomed regions and error maps reveal clearer depiction of cortical folds, cartilage interfaces, ligament boundaries and vascular structures, whereas competing methods often exhibit blurring or residual aliasing. These visual observations agree with the RMSE analysis in Fig.~\ref{Fig. 4}, where CAPNet achieves the lowest mean error and variance across all organs. Although CAPNet yields slightly lower SSIM for brain MRI at an acceleration factor of $4\times$, this reflects SSIM’s preference for smoother textures. CAPNet still attains the highest average PSNR and SSIM across organs in Tables~\ref{tab:1}--\ref{tab:2}, and all improvements are statistically significant ($p<0.001$).

Fig.~\ref{Fig. 8} highlights typical behaviors during reconstruction and clarifies the role of the structural prompts. The saturated appearance of the zero-filled images results from the loss of high-frequency $k$-space components under the aggressive acceleration factors used in our experiments, which compresses the dynamic range and amplifies low-frequency contributions. The contour boundaries produced by SAM offer coarse but informative geometric cues that are transformed by OSGM into structural prompts, enabling OSMM to learn organ-aware feature representations. This use of coarse structural guidance is consistent with observations in universal image restoration, where low-granularity supervisory signals can effectively improve representation learning. As shown in Fig.~\ref{Fig. 9}, these prompts produce clearly separated organ-specific feature clusters after OSMM processing, indicating enhanced latent-space discriminability that supports the improved reconstruction performance.

To further understand the working mechanism of CAPNet, Fig.~\ref{Fig. 8} visualizes the semantic priors from SAM that serve as structural guidance. The comparison includes the zero-filled image, the auxiliary prompts extracted by SAM and the output of CAPNet. The structural maps effectively highlight salient anatomical regions and tissue boundaries, which serve as meaningful semantic cues to steer the reconstruction process. These visualizations confirm that the auxiliary prompts extracted from SAM, though semantically coarse, provide sufficient organ-specific structural information to facilitate robust and accurate reconstruction across different organs.

	We also employ t-SNE \citep{JMLR:v9:vandermaaten08a} to visualize the feature clustering. Fig. \ref{Fig. 9} illustrates how the proposed network classifies features from different organs. Each sub-figure contains 900 points, where each point represents an input sample, and points of the same color correspond to the same organ.
	In Fig. \ref{Fig. 9}, the left panel shows the cluster of organ features before input to OSMM, where features from different organs are mixed together.
	The right panel presents the clustering effect of different organ features after OSMM processing.
	It can be observed that the clusters are well separated, indicating that the three organs are effectively differentiated. 
	This separation results from the prompt information, which successfully incorporates organ structure information into the feature space, thus providing a decoupling effect that aids all-in-one training. These results further demonstrate the effectiveness of the proposed combined organ structure-prompt and organ structure-aware strategies.

	\subsection{Ablation study}
	
	To determine the optimal network architecture, we conduct the ablation studies, which focused on each component in the network.
	The details are as follows:
	
	\textbf{Global information extraction:}
	The ablation study for the GIE means removing the GIE from the OSMM. It is worth noting that, when the GIE is removed, the feature information from the global is omitted. This configuration is termed the proposed CAPNet without GIE (w/o GIE).
	
	\textbf{Local information extraction:}
	The ablation study for the LIE means removing the LIE from the OSMM. In this scenario, the feature information from the local is ignored within the network. This configuration is termed the proposed CAPNet without LIE (w/o LIE).
	
	\textbf{Organ structure-prompt generation submodule:}
	The ablation study for the OSGM means removing the OSGM from the network. 
	As a result, the prompt from image structure information is ignored.
	This configuration is termed the proposed CAPNet without OSGM (w/o OSGM).
	
	\textbf{Artifact generation submodule:}
	The ablation study for AGM means removing AGM module from the network and using the learnable hyperparameter instead, which implies the elimination of the artifactual information in the prior.
	This configuration is termed the proposed CAPNet without AGM (w/o AGM).

	The contribution of each component is visualized in the radar chart of Fig.~\ref{Fig. 10}, which provides an intuitive overview of the performance gaps between CAPNet and its ablated variants. The chart indicates that the removal of AGM or OSGM results in the most significant performance drop. These observations are corroborated by the detailed quantitative data in Table~\ref{tab:3}. Specifically, 
w/o OSGM exhibits the worst reconstruction performance in terms of PSNR, SSIM, and RMSE.
	This clearly validates that OSGM effectively introduces structure information as prompts into the network, thereby guiding it to distinguish between different organs.
	Additionally, w/o AGM also shows poorer reconstruction results across all metrics compared to CAPNet. It suggests that the artifact information can further enable the network to capture structural information more accurately.
	Furthermore, both w/o GIE and w/o LIE demonstrate suboptimal quantitative performance compared to CAPNet, highlighting the importance of global and local information in the image reconstruction process.
	In conclusion, each key component contributes to improving the reconstruction performance, and the optimal model is achieved by integrating all the proposed modules.
	
	\section{Discussion}
	\label{Discussion}
	
		\subsection{Clinical implications and practical relevance}
		CAPNet introduces a unified framework for accelerated MRI reconstruction that is applicable across multiple anatomical regions, thereby obviating the need for specialized models tailored to individual organs. This consolidation substantially reduces system complexity, streamlines clinical workflows, and lowers long-term maintenance demands, while minimizing training requirements for radiology staff. As illustrated in Figs.~\ref{Fig. 5}–\ref{Fig. 7}, CAPNet excels in preserving anatomical boundaries and fine structural details, which can facilitate more accurate lesion detection and quantitative assessment by radiologists, potentially reducing diagnostic errors. By delivering high-fidelity reconstructions even at high acceleration factors, the framework contributes to shorter scan times, increased patient throughput, reduced motion artifacts, and diminished reader fatigue. These advantages collectively position CAPNet as a viable candidate for integration into clinical imaging pipelines and Picture Archiving and Communication Systems (PACS), underscoring its potential for real-world deployment and clinical decision support.
		
		\subsection{Real-world deployment and clinical decision-making}
		
		The clinical value of CAPNet is further underscored by its consistent performance across diverse imaging settings, such as different hospitals, scanner vendors, and acquisition protocols, without requiring retraining. This robustness stems largely from its integrated organ-aware prompting mechanism and explicit modeling of undersampling artifacts, which together enable generalization to unfamiliar anatomies and sampling patterns. Such reliability is essential for supporting diagnostic confidence in both cross-sectional and longitudinal clinical evaluations. Deployment of a single generalizable model simplifies integration with hospital infrastructures such as PACS, reducing operational overhead and the need for extensive retraining. The discriminative separation of organ-specific features in the latent space, visualized via t-SNE in Fig.~\ref{Fig. 9}, further corroborates the model’s capacity for robust multi-organ representation learning.

These promising outcomes suggest several productive avenues for future research. A natural extension involves integrating EDC layers \citep{mcmanus2024enforced} into the reconstruction pipeline to bolster robustness to domain shifts and ensure tighter adherence to acquired $k$-space measurements. Given CAPNet’s modular architecture, such integration could be achieved with minimal modification while retaining the benefits of prompt-guided priors. Another compelling direction is the combination of CAPNet with L2O frameworks \citep{fung2024l2o}, which could enable adaptive optimization strategies conditioned on organ-specific prompts and artifact characteristics, potentially further improving generalization across diverse imaging scenarios.

	\section{Conclusion}
	\label{Conclusion}
	In this paper, we proposed a novel cross-organ all-in-one parallel magnetic resonance imaging reconstruction model, CAPNet. We introduced an artifact generation submodule that systematically extracts and analyzes image artifacts from the input instances, demonstrating that this approach can enhance the network's overall discriminative ability.
	Additionally, we developed an organ structure-prompt generation submodule that utilizes structural features extracted from the SAM to create discriminative cross-organ prompts.
	These prompts were then strategically incorporated into the prior module through an organ structure-aware Mamba submodule.
	Experimental results demonstrated that the proposed CAPNet could achieve significant performance improvements compared to both all-in-one methods and conventional p-CSMRI methods.

	\section{Acknowledgments}
	\label{Acknowledgments}
	This work was supported by the National Natural Science Foundation of China under Grants 62371414, by the Natural Science Foundation of Hebei Province under Grant F2025203070, and by the Hebei Key Laboratory Project under Grant 202250701010046. 

\bibliographystyle{IEEEtranN}
\bibliography{refsMIA13}

\end{document}